\documentclass[aps,showpacs,prx,superscriptaddress,preprintnumbers,amsmath,amssymb,twocolumn,nofootinbib]{revtex4-2}

\usepackage{graphicx}%
\usepackage[usenames,dvipsnames]{xcolor}
\usepackage{siunitx}
\usepackage{subfigure}
\usepackage{enumitem}
\usepackage{subfigure}
\usepackage[export]{adjustbox}

\usepackage{tabularx}
\usepackage{booktabs}
\usepackage{multirow}

\usepackage{titlesec}

\usepackage{array}

\usepackage[most]{tcolorbox}

\usepackage[utf8]{inputenc}
\usepackage[T1]{fontenc}

\usepackage{bm}

\usepackage{lipsum}

\graphicspath{{Figures/}{}}

\usepackage[
bookmarks=true,
colorlinks,
linkcolor=black,
urlcolor=black,
citecolor=black,
plainpages=false,
pdfpagelabels,
final,
breaklinks=true
]{hyperref}

\hypersetup{
pdftitle={Decoding phase and time-dependent interferograms of high-oder harmonics}, 
pdfauthor={Camilo Granados, Ching-Ling Hsiao, Marcelo F Ciappina and Khadga Jung Karki}
}

\usepackage{natbib}
\makeatletter \def\NAT@def@citea{\def\@citea{\NAT@separator\,}} \makeatother

\newcommand{\orcid}[1]{%
  \href{%
    https://orcid.org/#1%
  }{%
   \,\protect\includegraphics[width=8pt]{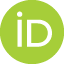}%
  }%
}

\newcommand{\rmd}{\ensuremath{\textrm{d}}}
\newcommand{\rmi}{\ensuremath{\textrm{i}}}

\usepackage{physics}

\DeclareSIUnit{\au}{{a.u.}}

\newcommand{\nhphantom}[1]{\sbox0{#1}\hspace{-\the\wd0}}

\begin{document}

\title{Decoding phase and time-dependent interferograms of high-order harmonics}

\author{C. Granados\,\orcid{0009-0001-9559-8978}}
\affiliation{Department of Physics, Guangdong Technion - Israel Institute of Technology, 241 Daxue Road, Shantou, Guangdong, China, 515063}
\affiliation{Technion -- Israel Institute of Technology, Haifa, 32000, Israel}
\affiliation{Guangdong Provincial Key Laboratory of Materials and Technologies for Energy Conversion, Guangdong Technion - Israel Institute of Technology, 241 Daxue Road, Shantou, Guangdong, China, 515063}

\author{Ching-Ling Hsiao}
\affiliation{Department of Physics, Guangdong Technion - Israel Institute of Technology, 241 Daxue Road, Shantou, Guangdong, China, 515063}
\affiliation{Technion -- Israel Institute of Technology, Haifa, 32000, Israel}
\affiliation{Guangdong Provincial Key Laboratory of Materials and Technologies for Energy Conversion, Guangdong Technion - Israel Institute of Technology, 241 Daxue Road, Shantou, Guangdong, China, 515063}
\affiliation{Department of Materials Science and Engineering, Guangdong Technion - Israel Institute of Technology, 241 Daxue Road, Shantou, Guangdong, China, 515063}

\author{Marcelo F. Ciappina\,\orcid{0000-0002-1123-6460}}
\email{marcelo.ciappina@gtiit.edu.cn}
\affiliation{Department of Physics, Guangdong Technion - Israel Institute of Technology, 241 Daxue Road, Shantou, Guangdong, China, 515063}
\affiliation{Technion -- Israel Institute of Technology, Haifa, 32000, Israel}
\affiliation{Guangdong Provincial Key Laboratory of Materials and Technologies for Energy Conversion, Guangdong Technion - Israel Institute of Technology, 241 Daxue Road, Shantou, Guangdong, China, 515063}

\author{Khadga J. Karki\,\orcid{0000-0002-0002-4163}}
\email{khadga.karki@gtiit.edu.cn}
\affiliation{Department of Physics, Guangdong Technion - Israel Institute of Technology, 241 Daxue Road, Shantou, Guangdong, China, 515063}
\affiliation{Technion -- Israel Institute of Technology, Haifa, 32000, Israel}
\affiliation{Guangdong Provincial Key Laboratory of Materials and Technologies for Energy Conversion, Guangdong Technion - Israel Institute of Technology, 241 Daxue Road, Shantou, Guangdong, China, 515063}

\date{\today}

\begin{abstract}
Interferometric measurements of high-harmonics induced by multiple laser fields in an emerging field of research that promises optimized yield of harmonics, and time and space-resolved nonlinear spectroscopy. Most of the measurements have been done by controlling the time-delay between the pulses. Here, we show that by changing one additional parameter, i.e. the phase-difference between the fields, together with the time-delay, one can, on the one hand, enhance the harmonic yield and, on the other hand, obtain in-depth information about the physical mechanisms that control the electron trajectories contributing to the high-harmonic generation. The two-dimensional interferograms obtained from such investigations can be used to find the values of time-delay and phase between the laser fields that maximize the yield of a particular harmonic. Results show that maximum yields of certain harmonics can be orders of magnitude larger than when using a single field or two fields with zero time-delay and phase difference. Our high-harmonics two-dimensional interferograms-based method paves the way for a simpler analysis of the attosecond electron dynamics in complex molecules and solids.  
\end{abstract}

\maketitle

\section{Introduction}

In strong-field laser-matter interaction, high-harmonic generation (HHG) is an extremely nonlinear optical process in which a strong laser field drives the emission of a train of short bursts of high-frequency (short wavelength) radiation~\cite{Krausz2009, Corkum2007}. This coherent radiation can cover hundreds of harmonic orders of the driving field, over a broad plateau that terminates at a sudden cutoff. The HHG phenomena is often explained through a semi-classical model that uses the so-called three-step process in which the laser ionizes the target atom via tunnel ionization and then propels the released electron back to the parent ion, where it recombines, releasing the kinetic energy acquired from the laser as a high-energy photon~\cite{Corkum1993, Kulander1993}.

Over the last three decades, substantial progress has been made in the understanding and potential applications of HHG \cite{RevModPhys.81.163,FemtoChem,HHGSolids,Khadga}. HHG offers the promise of creating new and reliable table-top coherent light sources from the extreme ultraviolet (XUV) to the x-ray spectral regions. Such new light sources can be produced, depending on the time duration of the driving strong laser field, (i) in the form of an isolated attosecond pulse (IAP), with a time duration as short as 100 as (1 as $=10^{-18}$ s) or less \cite{ShortestPulse}, or (ii) in the form of an attosecond pulse train (APT). The availability of attosecond pulses has boosted attosecond science in recent years in fields ranging from femtochemistry in gases to electron dynamics in liquid solutions and production of HHG is solid materials \cite{Liquids, chiral,correlations}. Another important application is HHG spectroscopy, which aims at obtaining atomic and molecular structural information from the experimentally measured HHG spectra. 
To advance further in the field, not only progress in technological and experimental tools but also theoretical insight are required. In this sense, theoretical simulations are instrumental in rationalizing the underlying mechanisms responsible for the observed features in the harmonics and decoding the information embedded within.

HHG in atoms can be modeled using a wide range of approaches, from classical heuristics~\cite{Corkum1993, Kulander1993} to intensive numerical computations~\cite{Scrinzi2014}, but the quantitative models that closely follow the overall intuition are quasiclassical methods~\cite{Lewenstein1994, Amini2019}. For example, the so-called Strong-Field Approximation~(SFA), where the emission amplitude is given by a path-integral summed over discrete emission events, is one of the most prominent and widely used theoretical method~\cite{Camilo,Amini2019}.
In SFA we can describe HHG in terms of quantum orbits~\cite{Salieres2001, Paulus2000, Kopold2002}, i.e.,~quasiclassical trajectories whose start and end times are complex quantities~\cite{Ivanov2014, Nayak2019}. It has also been demonstrated that dimensionally reduced models can be used to understand HHG. In this methods, for instance, one or two spatial coordinates are used to describe the active electron dynamics by taking advantage on the confined electron motion, along the laser electric field polarization axis.

Quantification and control of the electron trajectories is an emerging frontier that promises optimizing the yield of HHG and its technological applications in photolithography, optoelectronics, and time and space-resolved nonlinear spectroscopy~\cite{KONDO1996,Chipperfield2009,KIM2005,MAURITSSON2006,HUILLIER2013}. While HHG spectra in multi-pulse measurements have shown influence of multielectron dynamics in molecules and solids~\cite{IVANOV_2008,CORKUM_2015,DUDOVICH_2022}, their use in understanding the dynamical evolution of systems have seen limited progress. The main challenge has been in disentangling the pathways that contribute to the HHG spectra. In this regard, measurements done by varying the phase of one of the pulses may provide an additional degrees of freedom to disentangle the different pathways. While phase modulation techniques have been widely used in nonlinear spectroscopy in the perturbative regime of nonlinear interaction~\cite{MARCUS2007,KARKI2014C,WARREN_2005,KARKI_2019C}, their use in the nonperturbative interactions have not been explored yet. 

Recent developments have shown that it is feasible to simulate experiments done with multiple laser pulses `in-silico' using phase modulation~\cite{KARKI_2017C,KARKI_2019B,ANDA_2021,BRUSCHI_2022}. These methods are suitable not only for interpreting experimental results, but also in predicting the outcome of measurements thereby allowing one to assess the feasibility and benefits of committing resources for experiments. However, the methods developed so far have two drawbacks, i) they are based on the detection of population of excited states, and ii) the systems are parameterized based on a set of eigenstates rather than by realistic potentials. As a consequence, the simulations have been used only to investigate relatively slow dynamics induced by the coherent excitation of closely spaced energy levels.

In the present contribution, we implement a similar method to investigate ultrafast sub-cycle and cycle-averaged electron dynamics in a hydrogenic system induced by two phase-modulated strong laser fields. This is achieved by addressing the limitations of previous methods. Instead of detecting the excited states populations, we monitor the HHG spectra generated by the interaction between the strong laser field and the system. We model this interaction by the time-dependent Schr\"odinger equation of an electron in a hydrogenic potential perturbed by the strong field. This new method has allowed us to investigate different "fast" and "slow" processes that contribute to HHG and the dependence of the spectra on the phase-difference between the fields. This method can potentially be used to simulate electron/charge dynamics that contribute to HHG in more complex molecules and solids, and to interpret the complex spectra obtained in such systems. 

In the following sections, we describe the model system and the methods used in the simulations, present the results and discussion and finally give an outlook on the possible implementation of the method in actual experiments.

\section{Numerical Methods}
\label{numerics}%

In order to model the atomic-HHG we use a dimensionally-reduced time-dependent Schr\"odinger numerical solver. For linearly polarized laser pulses, the dynamics of an atomic electron in a strong laser field is mainly along the direction of the field polarization. It is reasonable, therefore, to model the HHG process in a 1D spatial dimension by solving the following Schr\"odinger equation (1D-TDSE):
\begin{eqnarray}
\label{tdse}
\rmi \frac{\partial \Psi(x,t)}{\partial t}&=&\mathcal{H}(t)\Psi(x,t) \\
&=&\left[-\frac{1}{2}\frac{\partial^{2}}{\partial x^{2}}+V_{\mathrm{atom}}(x)+V_{\mathrm{laser}}(x,t)\right]\Psi(x,t). \nonumber
\end{eqnarray}
To model an atom in 1D, it is common to use the quasi-Coulomb (or soft core) potential
\begin{eqnarray}
\label{atom}
V_{\mathrm{atom}}(x)&=&-\frac{1}{\sqrt{x^2+a^2}},
\end{eqnarray}
where the parameter $a$ is tuned to match the ionization potential $I_{p}$ of the atom under consideration. We use $a=1.41$ in order to model an electron in a 1s state of an H atom ($I_p=0.5$ a.u). The
potential due to the laser electric field linearly polarized along the $x$-axis is written in the dipole approximation as
\begin{eqnarray}
\label{vlaser}
V_{\mathrm{laser}}(x,t)&=&-E(\delta t,\phi, t)\,x,
\end{eqnarray}
where $E(\delta t,\phi,t)$ is the time-dependent laser electric field. Here, we employ a set of 2 identical pulses with a time delay $\delta t$ and a relative phase $\phi$ between them. Thus, we can write  $E(\delta t,\phi,t)$ as:
\begin{eqnarray}
\label{Efields}
E(\delta t,\phi,t)&=&E_1(0,t)+E_2(\phi,t-\delta t)
\end{eqnarray}
where each pulse is as follows ($j=1,2$):
\begin{eqnarray}
\label{sin2}
E_j(\phi,t)&=&E_0 \sin^{2}\left( \frac{\omega_L t}{2 N}\right)\sin(\omega_L t+\phi).
\end{eqnarray}
Here, $E_0$ is the laser electric field peak amplitude, $\omega_L$ is the laser carrier frequency, and $N$ is the number of total cycles.  The initial state in the 1D-TDSE is the ground state (GS) of the atomic system before we turn on the laser ($t=-\infty$) and it can be found by solving an eigenvalue problem once the spatial coordinate $x$ has been discretized in the space coordinate. 
Equation (\ref{tdse}) can be solved numerically by means of the Crank-Nicolson scheme~\cite{Ciappina2017}. Two important points should be highlighted: (i) to assure the stability of the numerical time evolution, the ratio between the resolution of time and space, i.e. $\delta t/\delta x$ should be sufficiently less than 1; (ii) in order to avoid reflections and spurious contributions to the electronic wavefunction a correct treatment of the boundary conditions is necessary, i.e.~the electronic wavefunction is multiplied by a mask function that is adequately tailored to \textit{absorb} such unwanted contributions.

Once having found the time-dependent electronic state $\Psi(x,t)$  from the 1D-TDSE (\ref{tdse}), the harmonic yield of an atom is proportional to the Fourier transform of the so-called time-dependent dipole acceleration $a(t)$ of its active electron. That is,
\begin{equation}
D(\omega)\propto \left|\frac{1}{\omega^2}\int_{-\infty}^{\infty}\rmd t\;\mathrm{e}^{\rmi \omega t}a(t)\right|^2,
\label{spec}
\end{equation}
where $a(t)$ can be obtained by using the following commutator relation:
\begin{equation}
a(t)=\frac{\rmd^{2}\langle x \rangle}{\rmd t^2}=-\langle \Psi(x,t) | \left[ \mathcal{H}(t),\left[ \mathcal{H}(t),x\right]\right] | \Psi(x,t) \rangle,
\end{equation}
where $\mathcal{H}(t)$ is the Hamiltonian defined in Eq.~(\ref{tdse}). The function $D(\omega)$
is called the dipole spectrum, since $D(\omega)$ gives the spectral profile measured in HHG experiments.
We drive the H atom by 2 identical pulses, with an intensity $I=1\times10^{14}$ W/cm$^2$ ($E_0=0.053$ a.u.) each, a wavelength $\lambda=800$ nm ($\omega_L=0.057$ a.u.) and 6 total cycles long ($\approx 15$ fs). Both the relative phase $\phi$ and time delay $\delta t$ are varied and, thus, $D(\omega)$ is now a function of these two parameters, $D(\omega,\phi,\delta t)$ .

\section{Results and discussion}
\label{results}

The variation of the relative phase and delay time between the two laser fields, in the HHG process, results in an interferogram with three distinctive features: i) The cut-off energy modulates with the phase and delay time,  ii) each individual harmonic shows fast sub-cycle modulations and iii) the harmonic yield is enhanced for different relative phase and delay time values. Figure 1(a) shows an interferogram of HHG spectra obtained by varying the phase of one of the pulses and keeping the time delay between the laser fields constant to 1.3 fs. The figure shows clearly the modulation of the cut-off energy as a function of the relative phase. Examples of the fast sub-cycle variation and the harmonic yield enhancement are shown in Figs.~1(b) and (c) for two particular harmonics. In the following we will analyze the cause of these features and their significance.    

\begin{figure}[h!]
 \centering{ \includegraphics[width=1 \linewidth, angle = -90]{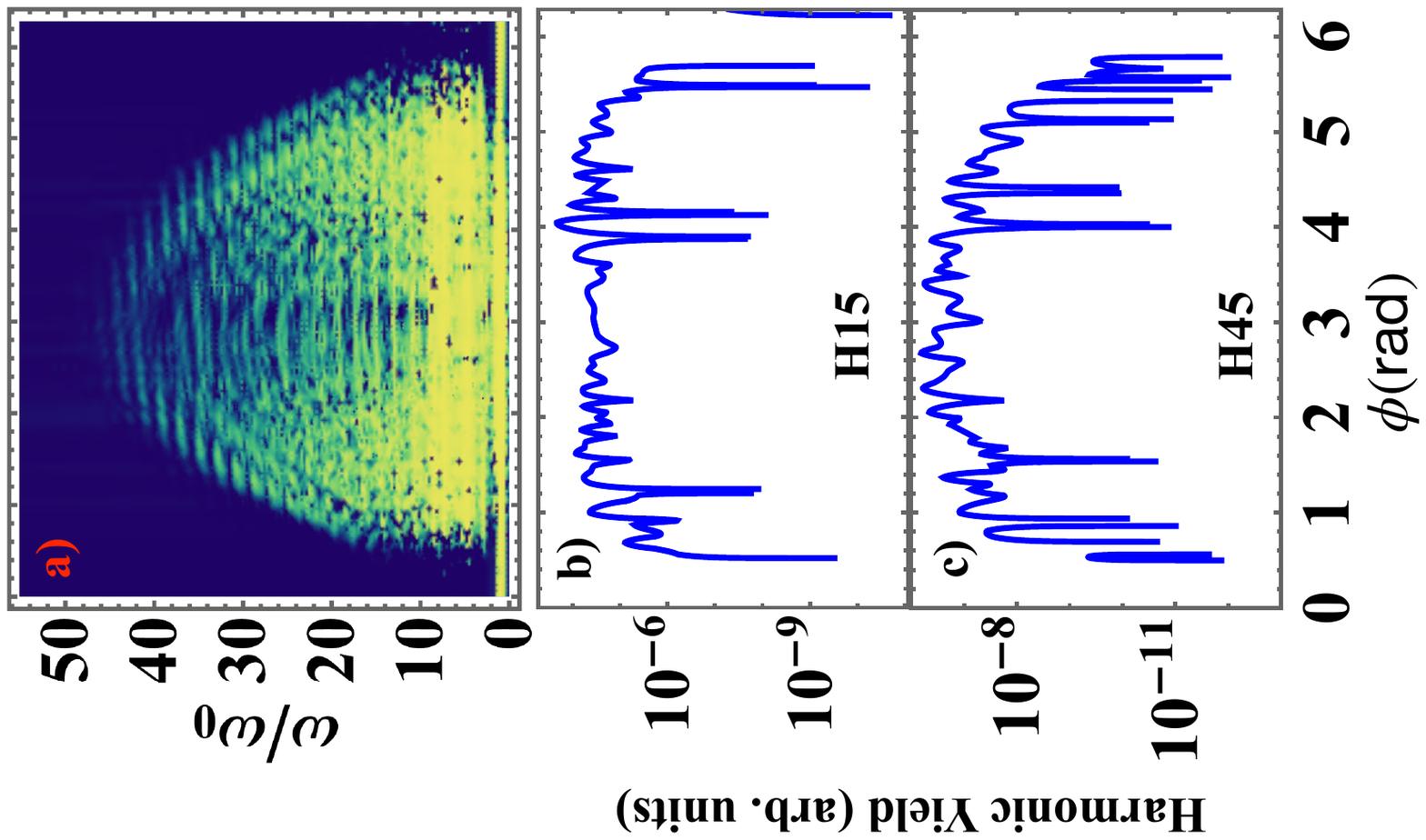}}
  \vspace{-5pt} 
  \caption{In a) density plot of the calculated HHG spectra extracted from the 1D-TDSE as a function of the phase difference. The delay value for this plot is set to $\delta t = 1.3$ fs. In b) and c) the yield for the 15$^\textrm{th}$ and 45$^\textrm{th}$ harmonics, respectively, as a function of the phase difference. Note that the dependence on the phase is not the same for these two harmonics.}
  \label{Results41}
\end{figure}

\subsection{Modulation of the cut-off}
\begin{figure}[h!]
\includegraphics[width=1.4 \linewidth, angle = -90]{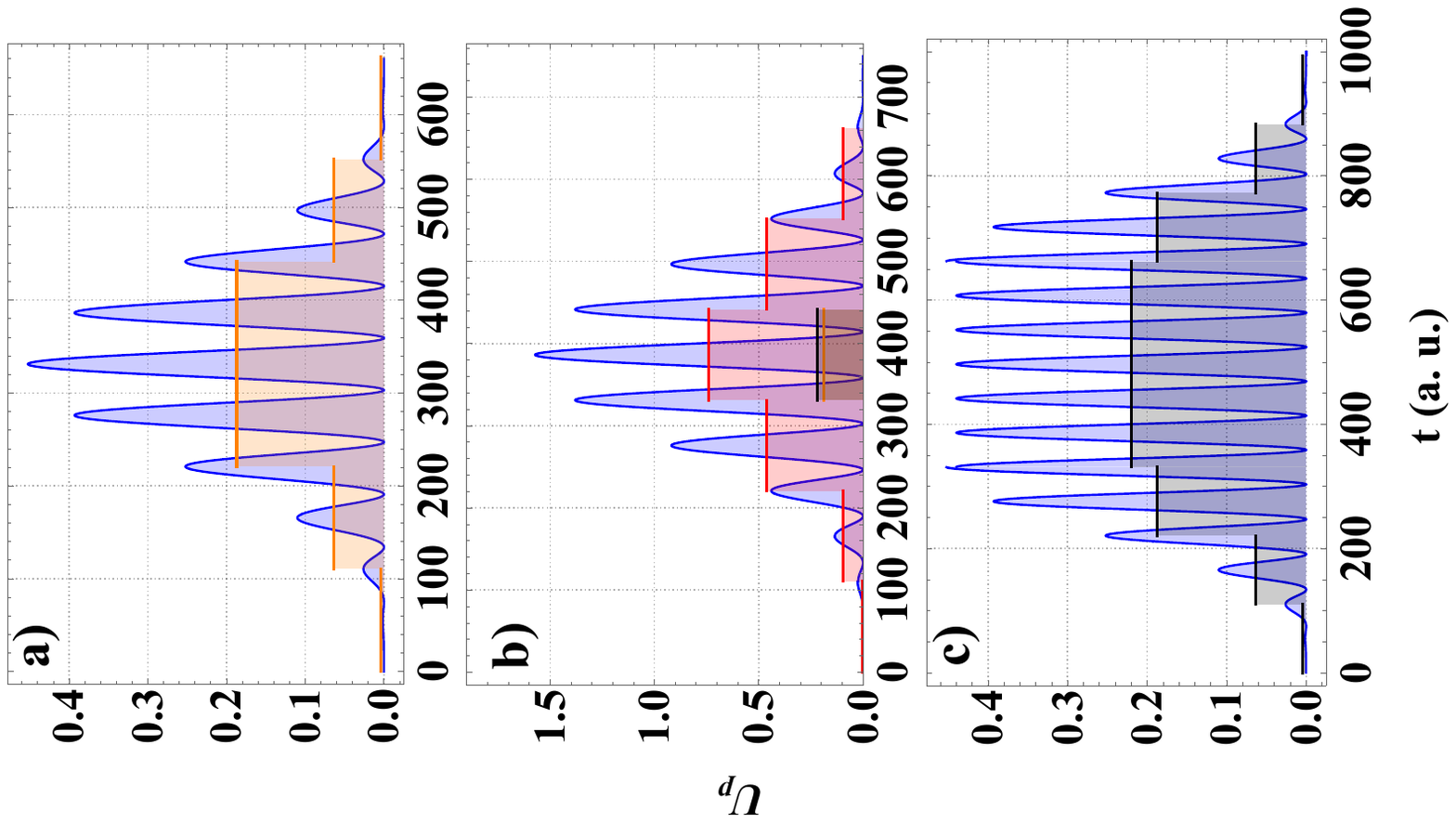}
\caption{Ponderomotive energy values for different pulse configurations. In solid blue line a) a single pulse $|A(t)|^2/2$, b) $|A_1(t)+A_2(t,0,T)|^2/2$, and c) $|A_1(t)+A_2(t,0,3T)|^2/2$. The maximum in the shadowed area in all the figures corresponds to the cycle-averaged $U_p(m)$ ($U_p(n,\phi,\delta t)$) calculated for each cycle. Additionally, we show in the middle panel a comparison between the maximum $U_p$ for the different cases.}
\label{SchematicUp}
\end{figure}

The calculation of the HHG spectra from the 1D-TDSE is a powerful method that, however, does not allow us to investigate analytically the HHG process and the physical mechanisms responsible for the observed experimental results. In particular, for this new scenario, where both the time delay and relative phase play a substantial role, we introduce a simple analysis of the modulation of the cut-off frequency based on the semiclassical properties of the HHG. It is well known that, for single pulse configurations, the maximum photon energy or the HHG cutoff, can be estimated by the well-known $3.17U_p$-law~\cite{Lewenstein1994, Amini2019}, where $U_p$ is the ponderomotive energy. The ponderomotive energy is defined as the cycle-averaged kinetic energy of the electron without considering its drift velocity, i.e.:

\begin{equation}
U_p =\dfrac{1}{2T} \int_{t}^{t+T}|A(t')|^2dt'
\label{Upmono}
\end{equation}

where $A(t)=-\int E(t') dt'$ is the laser vector potential and $T=2\pi/\omega_L$  is the oscillation period of the laser field. For a monochromatic wave (or a long pulse), where all the cycles are (almost) identical, the ponderomotive energy is given by the well known relation $U_p=E_0^2/4\omega_L^2$. However, for a few-cycle laser pulses, this definition should be revised (for more details see~\cite{Renata2016}). We can then define an optical cycle-dependent ponderomotive energy as~\cite{Renata2016}:
\begin{equation}
U_p (m)=\dfrac{1}{2T} \int_{(m-1) T}^{m T}|A(t')|^2dt',
\label{Up1pulse}
\end{equation}
with $m=1,\ldots,N$, $N$ being the total number of cycles. In our subsequent analysis, we neglect the contribution associated with the displacement vector $\bm{\alpha}(t)$ (see~\cite{Renata2016}).
The optical cycle-dependent ponderomotive energy $U_p (n,\phi,\delta t)$ for our two-pulse configuration can be written as:
\begin{equation}
U_p (n,\phi,\delta t) =\dfrac{1}{2T} \int_{(n-1)T}^{nT}|A(t',\phi,\delta)|^2dt', \label{UpEq}
\end{equation}
where now $n=1,2,\ldots,m+\lceil\frac{ \delta t }{T}\rceil$, $\lceil \cdot \rceil$ being the ceiling function, i.e.~the least integer greater than or equal to the argument. Equation~(\ref{UpEq}) involves the calculation of three different types of integrals, namely:

\begin{subequations}
\label{equations}
\begin{eqnarray}
 I_1(a_1,b_1) &=& \dfrac{1}{2T} \int_{a_1}^{b_1}|A_1(t)|^2dt,\label{eq:a}
 \\
  I_2(a_2,b_2) &=& \dfrac{1}{2T} \int_{a_2}^{b_2} |A_2(t,\phi,\delta t)|^2dt, \label{eq:b}\\
 I_3 (a_3,b_3)&=& \dfrac{1}{T} \int_{a_3}^{b_3}  |A_1(t)A_2(t,\phi,\delta t)|^2  dt \label{eq:c},
\end{eqnarray}
\end{subequations}
where the integration limits $[a_1,b_1], [a_2,b_2]$ and $[a_3,b_3]$ will be chosen accordingly to the value of $n$ and the time delay $\delta t$ between the two pulses. 
Each integral, $I_j(a_j,b_j)$, with $j=1,2,3$ describes different scenarios, described as follows: i) $I_1(a_1,b_1)$ corresponds to the ponderomotive energy $U_p$ for one single pulse, i.e.,  $A_1(t)$. ii) $I_2(a_2,b_2)$ is the contribution stemming only from the second field, i.e., $A_2(t,\phi,\delta t)$, and iii) $I_3(a_3,b_3)$ is the ponderomotive energy for the inter-pulse interference region. The analytical expressions for these integrals can be found in the Appendix. In order to show how to deal with the different cases, let us consider as an example a time delay $\delta t < T$. Thus, we can write the different $U_p (n,\phi,\delta t)$ values as:
\begin{eqnarray*}
U_p (1,\phi,\delta t)&=&I_1(0,\delta t)+I_3(\delta t,T)\\
U_p (j,\phi,\delta t)&=&I_3((j-1)T,jT)\\
U_p (7,\phi,\delta t)&=&I_2(6T,6T+\delta t),
   \label{Upex}
\end{eqnarray*}
with $j=2\ldots6$.
The resulting $U_p(n,\phi,\delta t)$ values from Eq.~(\ref{UpEq}) for three different cases are shown in Fig.~\ref{SchematicUp}, where we used $\phi = 0$ and the laser pulse parameters presented in Section II. In this figure, we compare the cases for a) one single pulse, b) two pulses delayed by $\delta t = T$, and c) two pulses delayed by $\delta t=3T$.  In Fig.~\ref{SchematicUp}(a) we can observe the usual behaviour for a single laser pulse~\cite{Renata2016}. On the other hand, for our two-pulse scheme, as shown in Figs.~\ref{SchematicUp}(b) and~\ref{SchematicUp}(c), we note that the maximum ponderomotive energy reaches a value approximately four times larger than the case of a single pulse, for a particular value of time delay $\delta t$ (we will also show that this value is a function of the relative phase $\phi$). This distinct behavior has profound consequences in the strong field electron dynamics initiated by the two-pulse excitation source.

\begin{figure}[h!]
 \centerline{ \includegraphics[width=1.05 \linewidth]{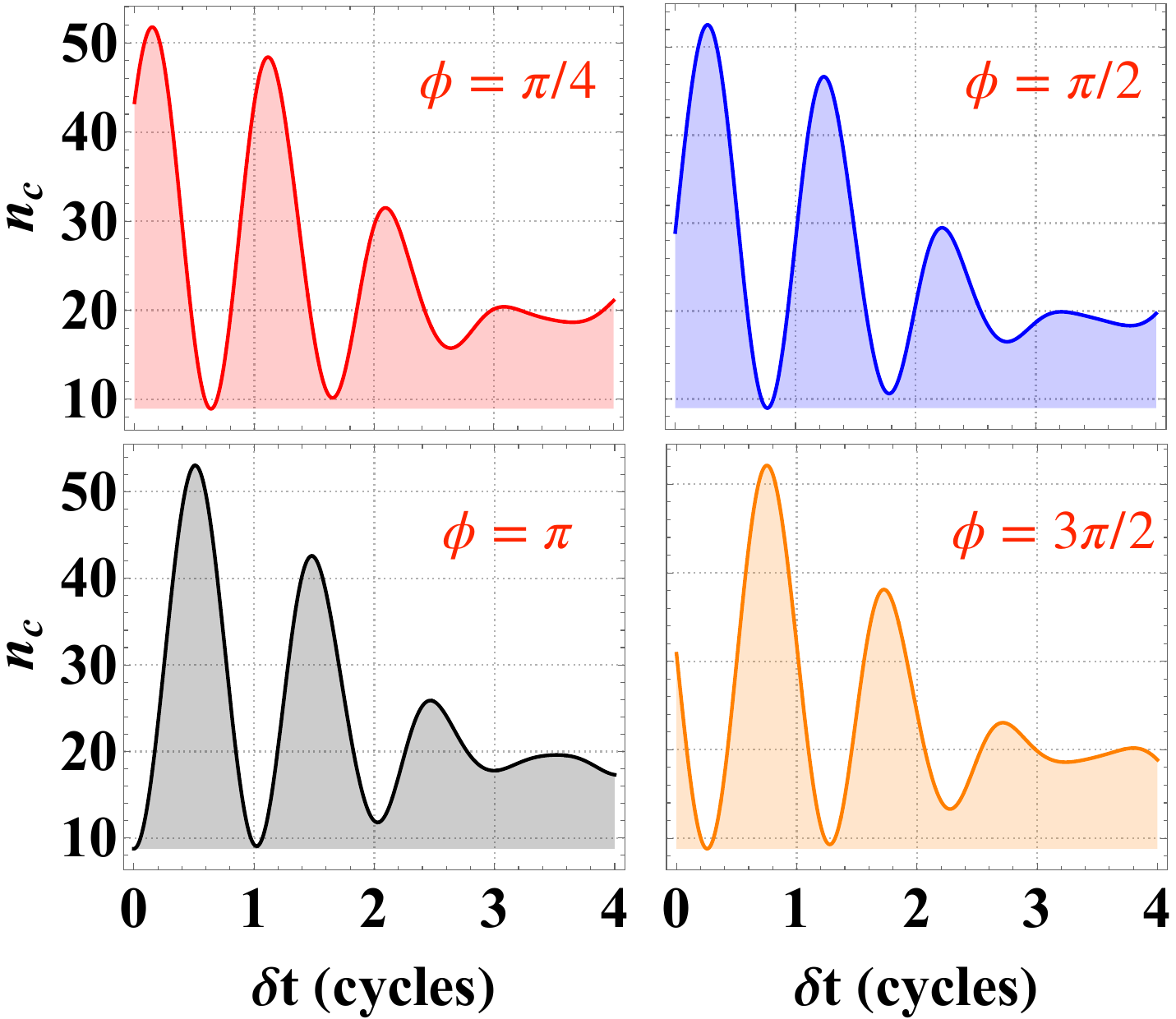}}
  \vspace{-5pt} 
  \caption{Harmonic cutoff $n_c$ values as a function of the delay $\delta t$ between the pulses and for different relative phases $\phi$. In all the panels we use $n = 4$ (see the text for more details).}
  \label{Results3}
\end{figure}

To understand the effect of the two-pulse configuration, we investigate the behavior of the harmonic cutoff, $n_c$. For this we computed $n_c$ from the semiclassical expression $n_c=(3.17 U_p^{max}+I_p)/\omega_L$ as a function of the time delay $\delta t$ for fixed phase values. By inspecting Fig.~\ref{SchematicUp}(b), we can observe that the maximum ponderomotive energy $U_p^{max}$ results for $n=4$. Thus, in Fig.~\ref{Results3} we plot the harmonic cutoff $n_c$ as a function of the time delay for different values of $\phi$. As $n_c$ scales linearly with $U_p^{max}$, it is not surprising to observe an increase when $U_p^{max}$ becomes larger. As demonstrated in Fig. \ref{Results3}, we observe a strong modulation of the harmonic cutoff, by changing both the time delay $\delta t$ and phase difference $\phi$ between the two pulses. Furthermore, Eq.~(\ref{UpEq}) allows us to study the influence of the interference term in our two-pulse scheme in an analytical fashion and find the values of $\delta t$ and $\phi$ that maximizes $n_c$. By evaluating Eq.~(\ref{UpEq}) for different values of $n$ we can obtain the delay $\delta t$ and phase difference $\phi$ values that maximize $U_p(n,\phi,\delta t$), and, therefore, $n_c$. We found a maximum value of $U_p^{max} = 0.8$ for $\phi = 2.9$ rad and $\delta t = 1.3$ fs. With the maximum ponderomotive energy, $U_p^{max}$, we can calculate the maximum harmonic cut-off,  $n_c = 53$. This value, for the present case, is $\approx 2$ times larger than the one for a single pulse. More importantly, the maximum cut-off is not at zero time delay and zero phase difference. In the following, we compare the analytical results with the simulations.

\subsection{1D-TDSE and comparison with the analytical solution}

\begin{figure}[h!]
\centerline{   \includegraphics[width=1.2 \linewidth]{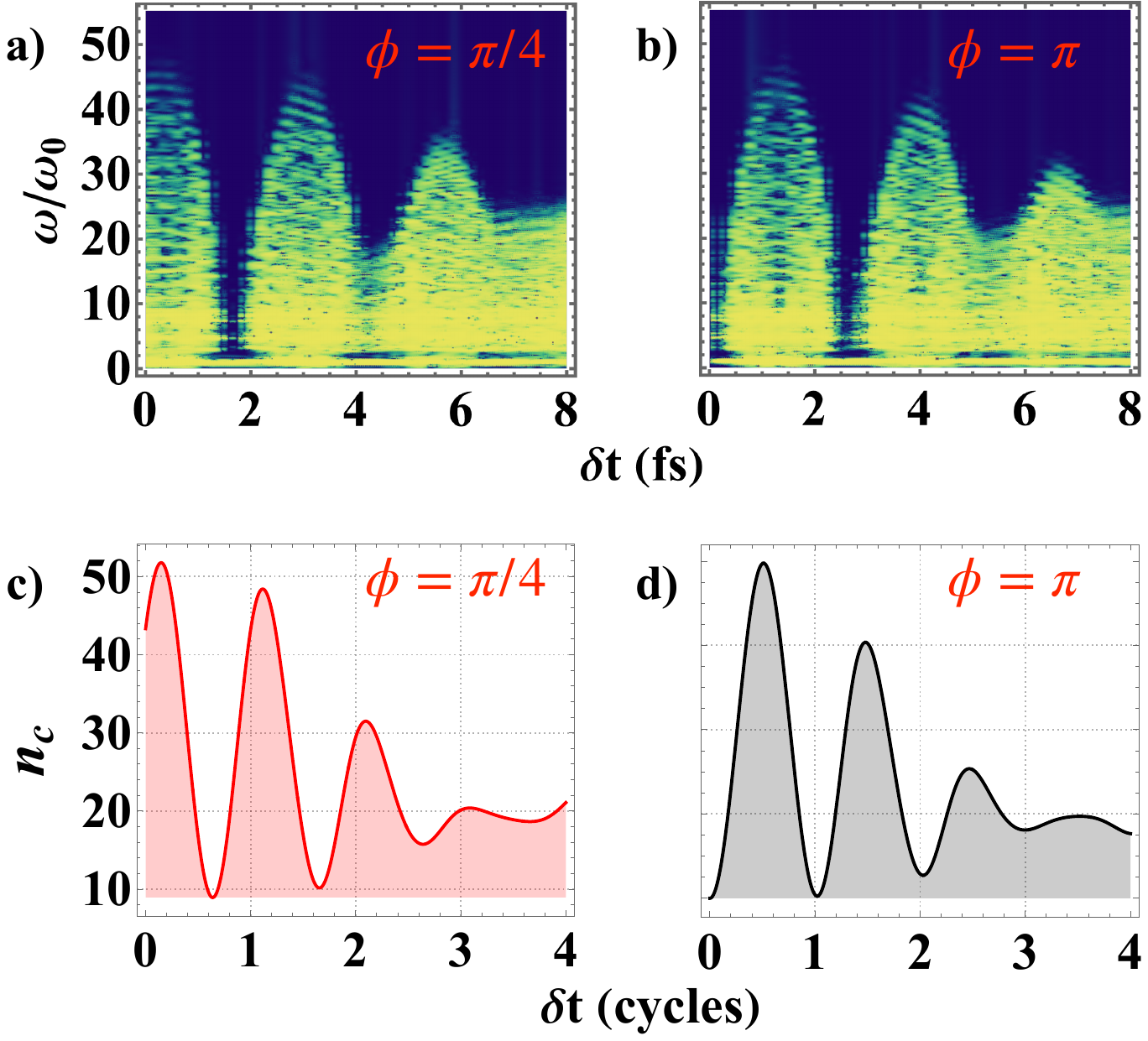}}
  \caption{(a) and (b) Density plot with the variation of the HHG spectra calculated from the 1D-TDSE as a function of the delay and for different values of the phase difference. The different cases shown in the panels corresponds to time delays in the interval $\{0, 15\}$ fs. (c) and (d): Calculated pulse averaged ponderomotive energy from Eq.~\ref{UpEq}. The results from the 1D-TDSE and the analytical calculations show that HHG cutoff is determined by the cycle averaged ponderomotive energy, and the cutoff is sensitive to the phase difference when the time delay between the pulses is less than three cycles of the field for pulses with a duration of six total cycles. }
  \label{Results21}
\end{figure}

One dimensional time dependent Schr\"odinger equation, 1D-TDSE, is used to calculate the effects of the time delay and phase difference between the two laser pulses on the harmonic spectra. In Figs. \ref{Results21} (a) and (b) we show HHG spectra as a function of time delay between the pulses at two values of phase differences ($\phi=\pi/2$ and $\phi=\pi$). Both spectra show modulations in the cutoff frequency with both the period of the optical cycle and the relative phases between the pulses. For pulses with six optical cycles, the modulations disappear after a time delay of about three optical cycles. These features are well reproduced by our analytical calculation based on the optical cycle-dependent ponderomotive energy, $U_p$, as shown in Fig. \ref{Results21} (c) and (d). 

\begin{figure}[h!]
\centerline{ \includegraphics[width=1.2 \linewidth,angle=-90]{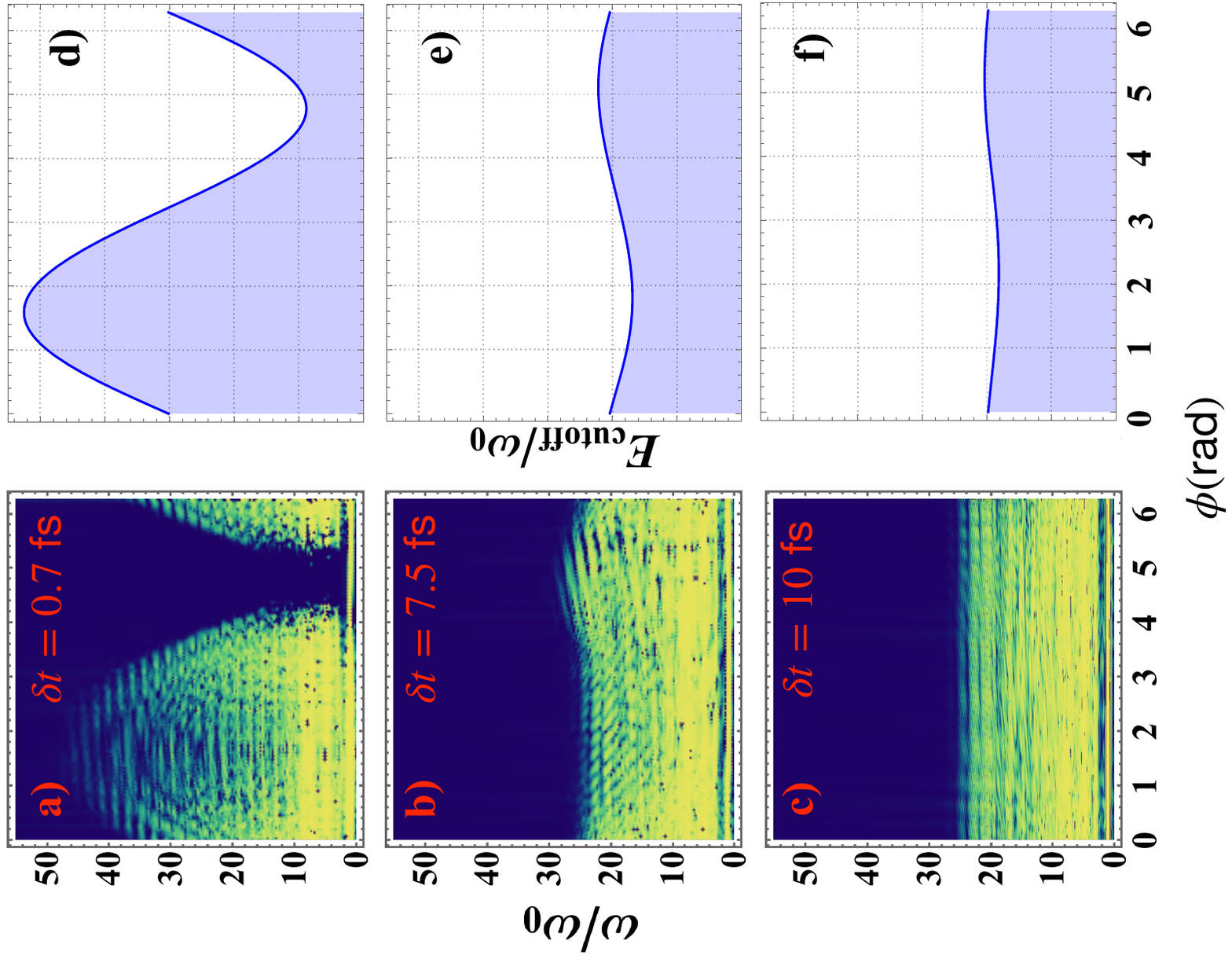}}
  \caption{(a,b,c) Density plots with the variation of the HHG spectra as a function of the phase difference at time delays $\delta t = 0.7$ fs, $7.5 $ fs and $10$ fs, respectively; (d,e,f): Equation~\ref{UpEq} plotted for the corresponding time delays. }
  \label{Results4}
\end{figure}

Calculation of the HHG spectra as a function of the phase difference for two different time delays, $\delta t=0.7$ fs, $7.5$ fs and $10$ fs are shown in Figs.~\ref{Results4} (a), (b) and (c). For short time delays, the cutoff frequency increase and decrease with a period of $2\pi$ in the phase difference. While the overall modulation of the cutoff frequency is obvious for $\delta t <2T$, it becomes less conspicuous for $2T<\delta<4T$ and negligible for $\delta t>4T$. These features are also reproduced well by our analytical calculations (Figs. \ref{Results4} (d), (e) and (f)), confirming that the cutoff frequency in HHG is given by the cycle dependent rather than the pulse averaged ponderomotive energy. The spectra obtained from the 1D-TDSE also show extremely fast modulations with both the phase difference (Fig.~\ref{Results4} (a) and (b)) and time delay (Fig. \ref{Results21} (a) and (b)). These features are not reproduced by the analytical calculations and deserve a more in-depth analysis. In the following section we investigate the HHG yield as a function of the two parameters: phase difference and time delay.

\subsection{HHG yield as a function of phase difference and time delay}
\begin{figure}[h!]
 \centerline{ \includegraphics[width=1.6 \linewidth, angle = -90]{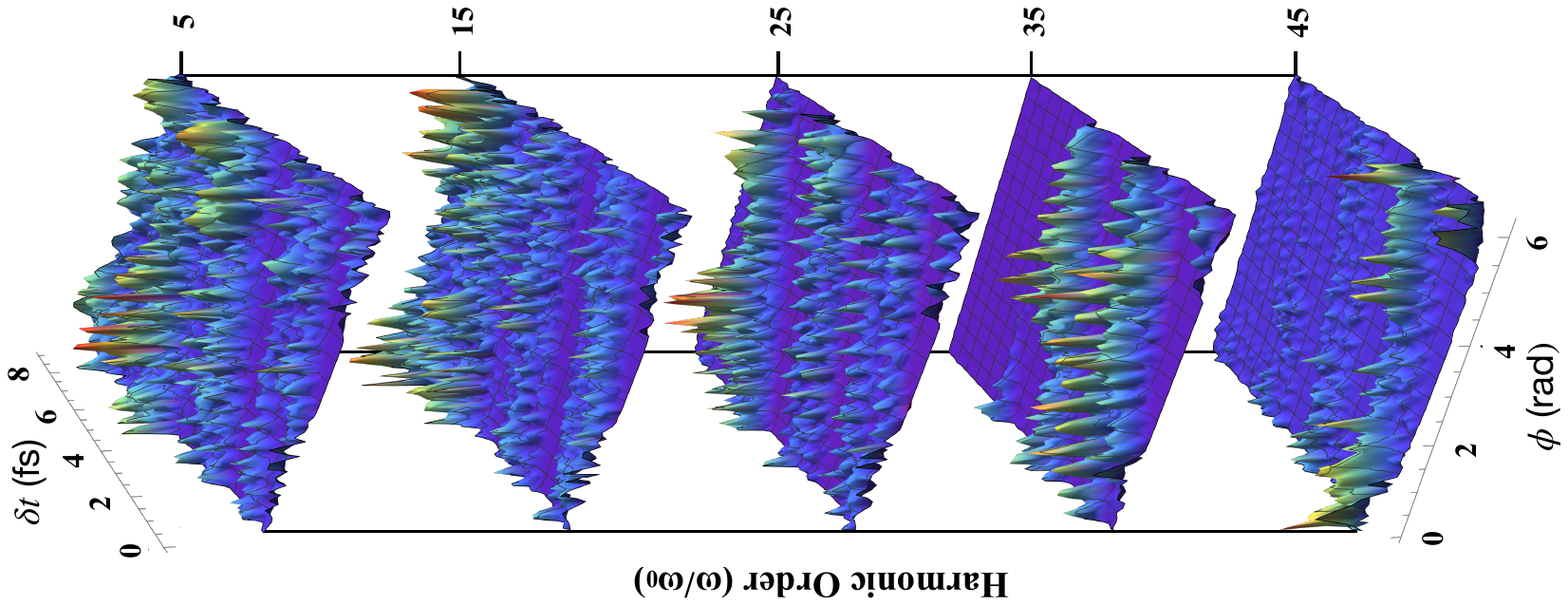}}
  \vspace{-5pt} 
  \caption{Three-dimensional interferograms with the dependence of the HHG yield for different harmonic orders on the phase difference and delay time.}
  \label{Results8}
\end{figure}

The observed modulations in the cut-off energy presented in Figs.~\ref{Results21} and \ref{Results4}, show that the dependence of the HHG yield on the phase difference and time delay is rather complex. HHG yields for selected harmonics (5$^\textrm{th}$, 15$^\textrm{th}$, 25$^\textrm{th}$, 35$^\textrm{th}$ and 45$^\textrm{th}$ ) are shown in in Fig.~\ref {Results8}. The sensitivity of the yields to both parameters is different for different harmonics. More importantly, not all the harmonics have a maximum at $\delta t=0$ and $\phi=0 $. For example, the 5$^\textrm{th}$ harmonic is maximum at $\phi\sim 3$ rad and $\delta t\sim6$ fs, the 25$^\textrm{th}$ at $\phi\sim4$ rad and $\delta t\sim7$ fs and the 45$^\textrm{th}$ at $\phi\sim6$ rad and $\delta t\sim0$ fs. In fact, for the same harmonics, the maximum harmonic yield is 9, 6 and 2 times larger than at zero time delay and phase difference. 

It is important to highlight that different methods have been proposed and implemented in order to increase the conversion efficiency~\cite{KONDO1996,Chipperfield2009,KIM2005,MAURITSSON2006,HUILLIER2013} using the interference between two pulses. However, in these methods, only the time delay between the pulses is varied to optimize the yield.  Our results show that a maximum harmonic yield cannot be obtained by varying the time delay alone. If the yield at particular harmonics is to be maximized, it appears necessary to use two beams where both the phase difference and time delay can be varied independently. This results is vital for applications in photolithography or generation of isolated attosecond pulses with high flux. 

The yield of any of the harmonics in Figs.~\ref{Results4}  and \ref{Results8} as a function of $\phi$ show some periodicity. Furthermore, the spectra show fast as well as slow modulations. Here, a phase difference of $\phi=2\pi$ corresponds to a period of the optical cycle $T_0$ and frequency $\omega_L$. For a particular case of an $m^{\textrm{th}}$ harmonic, one could consider two situations that lead to the modulations: (i) the pulses generate the coherent emissions independently at the frequency $m\omega_L$, whose interference modulates the yield with a period of $T_0/m)$ or (ii) the emissions arise from all possible combinations of the two fields that influence the electron trajectories. In the later case, the emission at $m\omega_L$ show modulations with multiple periods  given by $T_0/n$ , where $n$ is an integer, and it can be greater or smaller than $m$. 

\begin{figure}[h!]
 \centerline{ \includegraphics[width=1.3 \linewidth, angle = -90]{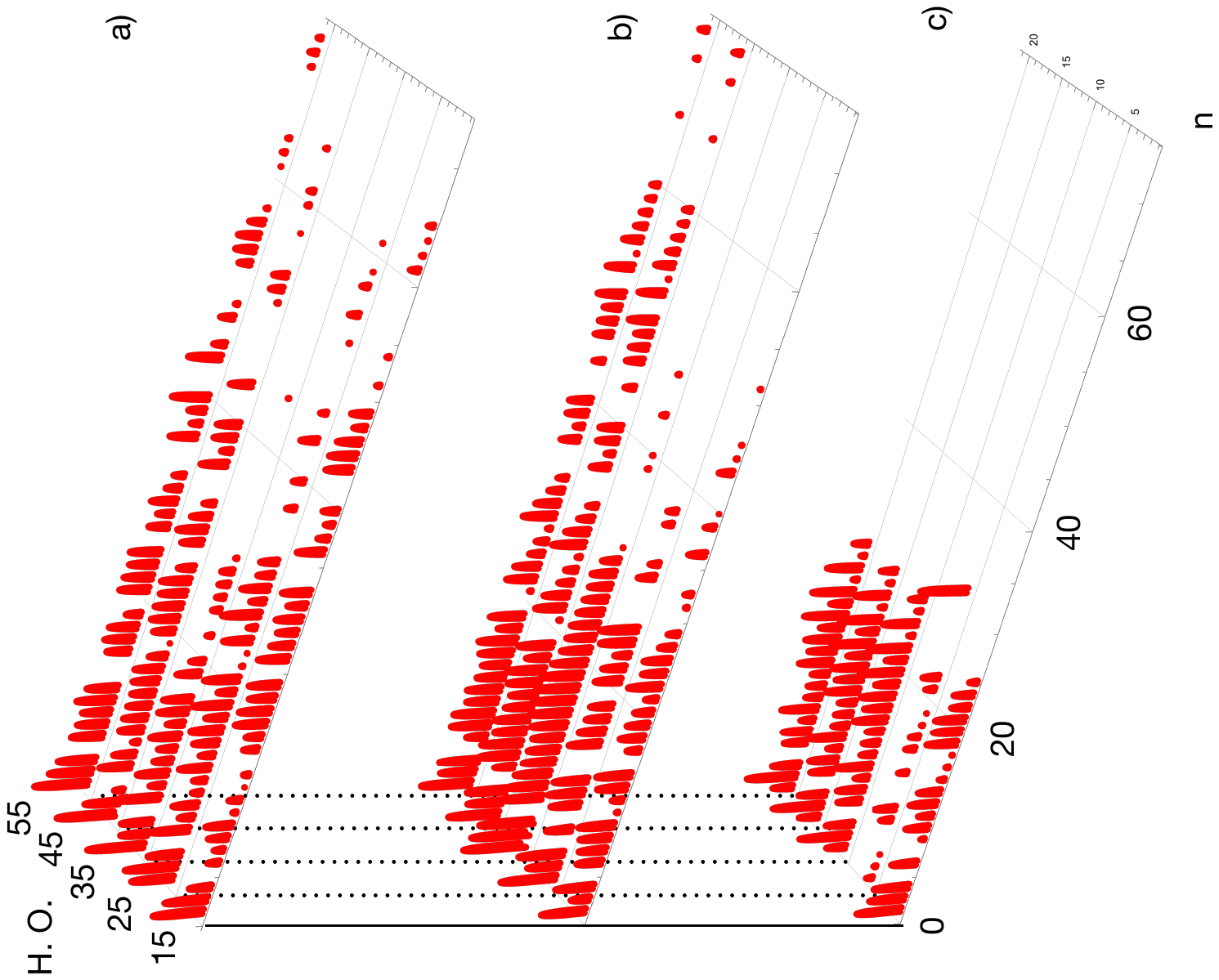}}
  \vspace{-5pt} 
  \caption{HHG dependence on the Fourier transformed phase. In the plots we use (a) $\delta t=$ 1.3 fs , (b) $\delta t=$ 5.5 fs, and (c) $\delta t=$ 10 fs. }
  \label{Results9}
\end{figure}

As shown in Fig.~ \ref{Results8}, the yield of any harmonic with respect to $\phi$ is modulated with different periods, indicating that the electron trajectories are determined by both pulses. We take the Fourier transform of the  harmonics to quantify the distribution of the trajectories of the electron that contribute to HHG, some of which are shown in Fig.~\ref{Results9}. The discrete peaks in the Fourier transformed data at $n/T_0$, allow us to quantify the trajectories as "fast" or "slow". The peaks with large values of $n$ in Figs.~\ref{Results9} (a) and (b) are from the fast trajectories with rapid phase evolution while the peaks with small values of $n$ are from the trajectories with slower phase evolution. The highest modulation frequency that can be observed in our simulations is with $n=70$, which corresponds to a time period of 35 as. As can be seen, fast as well as slow trajectories contribute to all the harmonics at small time delays, while only slow trajectories contribute at larger time delays, as shown in Fig.~\ref{Results9}(c). Moreover, the distribution for each of the harmonics evolves distinctly, indicating that it could be useful in tracing sub-cycle ultrafast dynamics induced by high fields with extremely high temporal resolution. 

The important results obtained from the simulations lead us to consider if it is feasible to control the phase difference in an experimental setup with such a high precision in order to quantify the distribution of the different electron trajectories imprinted in the HHG interferograms. In fact, in a typical measurement done with an amplified femtosecond laser at a repetition of few kHz, the measurements done by changing the phases would be too slow to overcome the smearing of the fast interferences due to different types of noises (vibrations, thermal drifts, electronic noise, etc.) that are prevalent at low frequencies. On the other hand, if high-repetition rate (megahertz) systems are used, one can cycle through the phase variations at kHz frequencies such that a lock-in detection can be used to precisely measure the amplitudes of the harmonics modulated at different frequencies. In this regard, high-repetition rate optical parametric chirped-pulse amplification systems, that have started to be commercially available, are ideally suitable. As the intensities that can be obtained from the megahertz systems are typically lower, HHG from molecules and solids, which require less intensity, could be investigated using the method. In fact, as HHG from molecules and solids would have a more complex spectra, an analysis based on phase modulation will aid in their interpretation.

\bigskip
\section{Conclusions}
To conclude, we have investigated the interferograms of HHG as a function of time delay and phase difference between two femtosecond pulses. The interferograms show modulations in the cut-off of HHG with the phase difference and each of the harmonics shows sub-cycle modulations with periods that are integer fractions of the cycle of the driving fields. Importantly, the maximum yield of the harmonics are sensitive to both the time-delay and phase difference, and for different harmonics they occur at different values. Lastly, the Fourier transforms of the interferograms as a function of phase difference of each of the harmonics shows a distinct evolution with the time-delay indicating that these spectra could be useful in disentangling the electron trajectories that contribute to the HHG, thereby aiding in the investigation of the relationship between the electronic structure and ultrafast sub-cycle electron dynamics.

\label{conclusions}

\acknowledgements

The present work is supported by the Guangdong Province Science and Technology Major Project (Future functional materials under extreme conditions - 2021B0301030005). M. F. C. and C. G. acknowledge financial support from the Guangdong Natural Science Foundation (General Program project No. 2023A1515010871).

\begin{widetext}
\section*{Appendix} \label{Appendix}

We include here the analytical expressions of the integrals $I_1(a_1,b_1)$, $I_2(a_2,b_2)$ and $I_3(a_3,b_3)$. The sin$^2$-shaped pulse of Eq.~(\ref{sin2}) can be written as~\cite{MilosevicReview}:
\begin{eqnarray}
E(t,\phi)&=&\sum_{i=0}^2\mathcal{E}_i\sin(\omega_i t+\phi),
\end{eqnarray}
where $\mathcal{E}_0=E_0/2$, $\mathcal{E}_1=\mathcal{E}_2=-\mathcal{E}_0/2$, $\omega_0=\omega_L$ and $\omega_{1,2}=\omega_L\pm\omega_p$, where $\omega_p=\omega_L/N$. Thus, the associated vector potential $A(t,\phi)=-\int_{\infty}^{t} E(t',\phi) \rmd t$ becomes
\begin{eqnarray}
A(t,\phi)&=&\sum_{i=0}^2 \mathcal{A}_i\cos(\omega_i t+\phi),
\end{eqnarray}
where $\mathcal{A}_i=\mathcal{E}_i/\omega_i$ ($i=0,1,2$). Using the above definition of $A(t,\phi)$ we can write the integrals Eqs.~(\ref{eq:a})-(\ref{eq:c}) as:

\begin{equation}
I_1(a_1,b_1)=  \dfrac{1}{2T}\int_{a_1}^{b_1} 
\Biggl[ \sum_{i=0}^{2} \Biggl( \mathcal{A}_i   \cos(\omega_i  t') \Biggr)^2 + 
2 \Biggl( \sum_{i=0}^{1} \sum_{k=i+1}^{2}\mathcal{A}_i\mathcal{A}_k\cos(\omega_i  t') \cos(\omega_k t') \Biggr)\Biggr] \rmd t' 
\end{equation}

\begin{eqnarray}
    I_2(a_2,b_2)&=&  \dfrac{1}{2T}\int_{a_2}^{b_2} 
\Biggl[ \sum_{i=0}^{2} \Biggl( \mathcal{A}_i   \cos(\omega_i  (t'-\delta t)+\phi) \Biggr)^2 \\
&+& 
2 \Biggl( \sum_{i=0}^{1} \sum_{k=i+1}^{2}\mathcal{A}_i\mathcal{A}_k\cos(\omega_i  (t'-\delta t)+\phi) \cos(\omega_k (t'-\delta t)+\phi) \Biggr)\Biggr] \rmd t' 
\end{eqnarray}

\begin{eqnarray}
I_3(a_3,b_3) &=& \dfrac{1}{T}\int_{a_3}^{b_3}  \Biggl[  \sum _{i=0}^1 \sum _{j=i+1}^2  \mathcal{A}_i\mathcal{A}_j\cos( \omega_i t')     \cos(\omega_j  (t' - \delta t)+\phi) \nonumber \\ &+&  \sum _{j=0}^1 \sum _{i=j+1}^2 \mathcal{A}_i\mathcal{A}_j \cos( \omega_i t')     \cos(\omega_j  (t' - \delta t) + \phi) \\
&+&  \sum _{i=0}^2   \mathcal{A}_i^2 \cos( \omega_i t') \cos (\omega_i  (t' - \delta t) + \phi) \Biggr] \rmd t'.
\nonumber
\end{eqnarray}

The explicit expressions for $I_1(a_1,b_1)$, $I_2(a_2,b_2)$ and $I_3(a_3,b_3)$ can be written as:

\begin{eqnarray}
I_1(a_1,b_1) &=&  \frac{1}{2 T} \Biggl[ \sum _{i=0}^2 \mathcal{A}_i^2 \Biggl(\frac{2 (b_1-a_1) \omega_i-\sin \left(2 a_1 \omega_i\right)+\sin \left(2 b_1 \omega_i\right)}{4 \omega_i}\Biggr) \nonumber \\
&+& 2\sum _{i=0}^1 \sum _{k=i+1}^2 \mathcal{A}_i\mathcal{A}_k\Biggl( \frac{\omega_i \left(\sin \left(b_1 \omega_i\right) \cos \left(b_1 \omega_k\right)-\sin \left(a_1 \omega_i\right) \cos \left(a_1 \omega_k\right)\right)}{\left(\omega_i^2-\omega_k^2\right)} \nonumber \\
&+& \frac{\omega_k \left(\cos \left(a_1 \omega_i\right) \sin \left(a_1 \omega_k\right)-\cos \left(b_1 \omega_i\right) \sin \left(b_1 \omega_k\right)\right)}{(\omega_i^2-\omega_k^2)} \Biggr)\Biggr],\label{I1}
\end{eqnarray}

\begin{eqnarray}
I_2(a_2,b_2) &=&  \frac{1}{2 T} \Biggl[ \sum _{i=0}^2 \mathcal{A}_i^2 \Biggl(\frac{2 (b_2-a_2) \omega_i-\sin \left[2(( a_2-\delta t) \omega_i+\phi)\right]+\sin \left[2(( b_2-\delta t) \omega_i+\phi)\right]}{\left(4 \omega_i\right)}\Biggr) \nonumber \\
&+& 2\sum _{i=0}^1 \sum _{k=i+1}^2 \mathcal{A}_i\mathcal{A}_k \Biggl( \frac{\sin \left[(b_2-\delta t ) \left(\omega_i-\omega_k\right)\right]-\sin \left[(a_2-\delta t ) \left(\omega_i-\omega_k\right)\right]}{(\omega_i-\omega_k)}  \nonumber \\
&+& \frac{\sin \left[(b_2-\delta t) \left(\omega_i+\omega_k\right)+2 \phi \right]-\sin \left[(a_2-\delta t) \left(\omega_i+\omega_k\right)+2 \phi \right]}{(\omega_i+\omega_k)} \Biggr)\Biggr]\label{I2}
\end{eqnarray}
and

\begin{eqnarray}
I_3(a_3,b_3)&=& \frac{1}{T} \Biggl[\sum _{i=0}^2 \mathcal{A}_i^2 \Biggl( \frac{2 (b_3-a_3) \omega_i \cos \left(\phi -\omega_i\delta t\right)-\sin \left[(2 a_3-\delta t) \omega_i+\phi \right]+\sin \left[(2 b_3-\delta t ) \omega_i+\phi \right]}{4 \omega_i} \Biggr) \nonumber \\
&+&  \sum_{i=0}^1 \sum _{j=i+1}^2 \mathcal{A}_i\mathcal{A}_j \Biggl( \frac{\omega_i \left(2 \sin \left(b_3 \omega_i\right) \cos \left[(b_3-\delta t ) \omega_j+\phi \right]-2 \sin \left(a_3 \omega_i\right) \cos \left[(a_3-\delta t ) \omega_j+\phi \right]\right)}{2 \left(\omega_i^2-\omega_j^2\right)} \nonumber \\
&+&\frac{\omega_j(\sin(a_3 \omega_i+(a_3-\delta t ) \omega_j+\phi )+\sin(-a_3 \omega_i+(a_3-\delta t ) \omega_j+\phi ))}{2(\omega_i^2-\omega_j^2)}\nonumber\\ 
&-& \frac{2\omega_j (\cos \left(b_3 \omega_i\right) \sin \left((b_3-\delta t ) \omega_j+\phi \right))}{2 \left(\omega_i^2-\omega_j^2\right)}\nonumber
\Biggr) \nonumber \\ 
&+& \sum _{j=0}^1 \sum _{i=j+1}^2 \mathcal{A}_i\mathcal{A}_j \Biggl( \frac{\omega_i \left(2 \sin \left(b_3 \omega_i\right) \cos \left[(b_3-\delta t ) \omega_j+\phi \right]-2\sin \left(a_3 \omega_i\right) \cos \left[(a_3-\delta  t) \omega_j+\phi \right]\right)}{2 \left(\omega_i^2-\omega_j^2\right)} \nonumber \\
&+& \frac{\omega_j \left(\sin (a_3 \omega_i+(a_3-\delta t ) \omega_j+\phi )+\sin (-a_3 \omega_i+(a_3-\delta t ) \omega_j+\phi)\right)}\nonumber \\
&-&\frac{2 \omega_j (\cos (b_3 \omega_i) \sin [(b_3-\delta t ) \omega_j+\phi ])}{2 (\omega_i^2-\omega_j^2)} \Biggr)\Biggr].\label{I3}
\end{eqnarray}

\end{widetext}

\bibliography{references,res_phase_mod}{}

\begin{thebibliography}{40}%
\makeatletter
\providecommand \@ifxundefined [1]{%
 \@ifx{#1\undefined}
}%
\providecommand \@ifnum [1]{%
 \ifnum #1\expandafter \@firstoftwo
 \else \expandafter \@secondoftwo
 \fi
}%
\providecommand \@ifx [1]{%
 \ifx #1\expandafter \@firstoftwo
 \else \expandafter \@secondoftwo
 \fi
}%
\providecommand \natexlab [1]{#1}%
\providecommand \enquote  [1]{``#1''}%
\providecommand \bibnamefont  [1]{#1}%
\providecommand \bibfnamefont [1]{#1}%
\providecommand \citenamefont [1]{#1}%
\providecommand \href@noop [0]{\@secondoftwo}%
\providecommand \href [0]{\begingroup \@sanitize@url \@href}%
\providecommand \@href[1]{\@@startlink{#1}\@@href}%
\providecommand \@@href[1]{\endgroup#1\@@endlink}%
\providecommand \@sanitize@url [0]{\catcode `\\12\catcode `\$12\catcode
  `\&12\catcode `\#12\catcode `\^12\catcode `\_12\catcode `\%12\relax}%
\providecommand \@@startlink[1]{}%
\providecommand \@@endlink[0]{}%
\providecommand \url  [0]{\begingroup\@sanitize@url \@url }%
\providecommand \@url [1]{\endgroup\@href {#1}{\urlprefix }}%
\providecommand \urlprefix  [0]{URL }%
\providecommand \Eprint [0]{\href }%
\providecommand \doibase [0]{https://doi.org/}%
\providecommand \selectlanguage [0]{\@gobble}%
\providecommand \bibinfo  [0]{\@secondoftwo}%
\providecommand \bibfield  [0]{\@secondoftwo}%
\providecommand \translation [1]{[#1]}%
\providecommand \BibitemOpen [0]{}%
\providecommand \bibitemStop [0]{}%
\providecommand \bibitemNoStop [0]{.\EOS\space}%
\providecommand \EOS [0]{\spacefactor3000\relax}%
\providecommand \BibitemShut  [1]{\csname bibitem#1\endcsname}%
\let\auto@bib@innerbib\@empty
\bibitem [{\citenamefont {Krausz}\ and\ \citenamefont
  {Ivanov}(2009{\natexlab{a}})}]{Krausz2009}%
  \BibitemOpen
  \bibfield  {author} {\bibinfo {author} {\bibfnamefont {F.}~\bibnamefont
  {Krausz}}\ and\ \bibinfo {author} {\bibfnamefont {M.}~\bibnamefont
  {Ivanov}},\ }\bibfield  {title} {\bibinfo {title} {Attosecond physics},\
  }\href {https://doi.org/10.1103/RevModPhys.81.163} {\bibfield  {journal}
  {\bibinfo  {journal} {Rev. Mod. Phys.}\ }\textbf {\bibinfo {volume} {81}},\
  \bibinfo {pages} {163} (\bibinfo {year} {2009}{\natexlab{a}})}\BibitemShut
  {NoStop}%
\bibitem [{\citenamefont {Corkum}\ and\ \citenamefont
  {Krausz}(2007)}]{Corkum2007}%
  \BibitemOpen
  \bibfield  {author} {\bibinfo {author} {\bibfnamefont {P.~B.}\ \bibnamefont
  {Corkum}}\ and\ \bibinfo {author} {\bibfnamefont {F.}~\bibnamefont
  {Krausz}},\ }\bibfield  {title} {\bibinfo {title} {Attosecond science},\
  }\href {https://doi.org/10.1038/nphys620} {\bibfield  {journal} {\bibinfo
  {journal} {Nat. Phys.}\ }\textbf {\bibinfo {volume} {3}},\ \bibinfo {pages}
  {381} (\bibinfo {year} {2007})}\BibitemShut {NoStop}%
\bibitem [{\citenamefont {Corkum}(1993)}]{Corkum1993}%
  \BibitemOpen
  \bibfield  {author} {\bibinfo {author} {\bibfnamefont {P.~B.}\ \bibnamefont
  {Corkum}},\ }\bibfield  {title} {\bibinfo {title} {Plasma perspective on
  strong field multiphoton ionization},\ }\href
  {https://doi.org/10.1103/PhysRevLett.71.1994} {\bibfield  {journal} {\bibinfo
   {journal} {Phys. Rev. Lett.}\ }\textbf {\bibinfo {volume} {71}},\ \bibinfo
  {pages} {1994} (\bibinfo {year} {1993})}\BibitemShut {NoStop}%
\bibitem [{\citenamefont {Kulander}\ \emph {et~al.}(1993)\citenamefont
  {Kulander}, \citenamefont {Schafer},\ and\ \citenamefont
  {Krause}}]{Kulander1993}%
  \BibitemOpen
  \bibfield  {author} {\bibinfo {author} {\bibfnamefont {K.~C.}\ \bibnamefont
  {Kulander}}, \bibinfo {author} {\bibfnamefont {K.~J.}\ \bibnamefont
  {Schafer}},\ and\ \bibinfo {author} {\bibfnamefont {J.~L.}\ \bibnamefont
  {Krause}},\ }\bibfield  {title} {\bibinfo {title} {Dynamics of short-pulse
  excitation, ionization and harmonic conversion},\ }in\ \href@noop {} {\emph
  {\bibinfo {booktitle}
  {\href{https://www.springer.com/gp/book/9780306445873}{Super-Intense Laser
  Atom Physics}}}},\ \bibinfo {series} {NATO Advanced Studies Institute Series
  B: Physics}, Vol.\ \bibinfo {volume} {316},\ \bibinfo {editor} {edited by\
  \bibinfo {editor} {\bibfnamefont {B.}~\bibnamefont {Piraux}}, \bibinfo
  {editor} {\bibfnamefont {A.}~\bibnamefont {L'Huillier}},\ and\ \bibinfo
  {editor} {\bibfnamefont {K.}~\bibnamefont {Rz\k{a}\.zewski}}}\ (\bibinfo
  {publisher} {Plenum},\ \bibinfo {address} {New York},\ \bibinfo {year}
  {1993})\ pp.\ \bibinfo {pages} {95--110}\BibitemShut {NoStop}%
\bibitem [{\citenamefont {Krausz}\ and\ \citenamefont
  {Ivanov}(2009{\natexlab{b}})}]{RevModPhys.81.163}%
  \BibitemOpen
  \bibfield  {author} {\bibinfo {author} {\bibfnamefont {F.}~\bibnamefont
  {Krausz}}\ and\ \bibinfo {author} {\bibfnamefont {M.}~\bibnamefont
  {Ivanov}},\ }\bibfield  {title} {\bibinfo {title} {Attosecond physics},\
  }\href {https://doi.org/10.1103/RevModPhys.81.163} {\bibfield  {journal}
  {\bibinfo  {journal} {Rev. Mod. Phys.}\ }\textbf {\bibinfo {volume} {81}},\
  \bibinfo {pages} {163} (\bibinfo {year} {2009}{\natexlab{b}})}\BibitemShut
  {NoStop}%
\bibitem [{\citenamefont {Zewail}(2000)}]{FemtoChem}%
  \BibitemOpen
  \bibfield  {author} {\bibinfo {author} {\bibfnamefont {A.~H.}\ \bibnamefont
  {Zewail}},\ }\bibfield  {title} {\bibinfo {title} {Femtochemistry:
  Atomic-scale dynamics of the chemical bond},\ }\href@noop {} {\bibfield
  {journal} {\bibinfo  {journal} {J. Phys. Chem. A}\ }\textbf {\bibinfo
  {volume} {104}} (\bibinfo {year} {2000})}\BibitemShut {NoStop}%
\bibitem [{\citenamefont {Ghimire}\ and\ \citenamefont
  {Reis}(2019)}]{HHGSolids}%
  \BibitemOpen
  \bibfield  {author} {\bibinfo {author} {\bibfnamefont {S.}~\bibnamefont
  {Ghimire}}\ and\ \bibinfo {author} {\bibfnamefont {D.}~\bibnamefont {Reis}},\
  }\bibfield  {title} {\bibinfo {title} {High-harmonic generation from
  solids},\ }\href {https://doi.org/10.1038/s41567-018-0315-5} {\bibfield
  {journal} {\bibinfo  {journal} {Nature Phys}\ }\textbf {\bibinfo {volume}
  {15}},\ \bibinfo {pages} {10} (\bibinfo {year} {2019})}\BibitemShut {NoStop}%
\bibitem [{\citenamefont {Karki}\ and\ \citenamefont
  {Ciappina}(2022)}]{Khadga}%
  \BibitemOpen
  \bibfield  {author} {\bibinfo {author} {\bibfnamefont {K.~J.}\ \bibnamefont
  {Karki}}\ and\ \bibinfo {author} {\bibfnamefont {M.~F.}\ \bibnamefont
  {Ciappina}},\ }\bibfield  {title} {\bibinfo {title} {Advances in nonlinear
  spectroscopy using phase modulated light fields: prospective applications in
  perturbative and non-perturbative regimes},\ }\href
  {https://doi.org/10.1080/23746149.2022.2090856} {\bibfield  {journal}
  {\bibinfo  {journal} {Advances in Physics: X}\ }\textbf {\bibinfo {volume}
  {7}},\ \bibinfo {pages} {2090856} (\bibinfo {year} {2022})}\BibitemShut
  {NoStop}%
\bibitem [{\citenamefont {Kim~H.Y.}()}]{ShortestPulse}%
  \BibitemOpen
  \bibfield  {author} {\bibinfo {author} {\bibfnamefont {M.~S. e.~a.}\
  \bibnamefont {Kim~H.Y.}, \bibfnamefont {Garg~M.}},\ }\bibfield  {title}
  {\bibinfo {title} {Attosecond field emission.},\ }\bibfield  {journal}
  {\bibinfo  {journal} {Nature 613, 662–666 (2023).}\ }\href
  {https://doi.org/10.1038/s41586-022-05577-1}
  {10.1038/s41586-022-05577-1}\BibitemShut {NoStop}%
\bibitem [{\citenamefont {Ikonnikov}\ \emph {et~al.}(2023)\citenamefont
  {Ikonnikov}, \citenamefont {Paolino}, \citenamefont {Garcia-Alvarez},
  \citenamefont {Orozco-Gonzalez}, \citenamefont {Granados}, \citenamefont
  {Röder}, \citenamefont {Léonard}, \citenamefont {Olivucci}, \citenamefont
  {Haacke}, \citenamefont {Kornilov},\ and\ \citenamefont {Gozem}}]{Liquids}%
  \BibitemOpen
  \bibfield  {author} {\bibinfo {author} {\bibfnamefont {E.}~\bibnamefont
  {Ikonnikov}}, \bibinfo {author} {\bibfnamefont {M.}~\bibnamefont {Paolino}},
  \bibinfo {author} {\bibfnamefont {J.~C.}\ \bibnamefont {Garcia-Alvarez}},
  \bibinfo {author} {\bibfnamefont {Y.}~\bibnamefont {Orozco-Gonzalez}},
  \bibinfo {author} {\bibfnamefont {C.}~\bibnamefont {Granados}}, \bibinfo
  {author} {\bibfnamefont {A.}~\bibnamefont {Röder}}, \bibinfo {author}
  {\bibfnamefont {J.}~\bibnamefont {Léonard}}, \bibinfo {author}
  {\bibfnamefont {M.}~\bibnamefont {Olivucci}}, \bibinfo {author}
  {\bibfnamefont {S.}~\bibnamefont {Haacke}}, \bibinfo {author} {\bibfnamefont
  {O.}~\bibnamefont {Kornilov}},\ and\ \bibinfo {author} {\bibfnamefont
  {S.}~\bibnamefont {Gozem}},\ }\bibfield  {title} {\bibinfo {title}
  {Photoelectron spectroscopy of oppositely charged molecular switches in the
  aqueous phase: Theory and experiment},\ }\href
  {https://doi.org/10.1021/acs.jpclett.3c00828} {\bibfield  {journal} {\bibinfo
   {journal} {The Journal of Physical Chemistry Letters}\ }\textbf {\bibinfo
  {volume} {14}},\ \bibinfo {pages} {6061} (\bibinfo {year} {2023})},\ \bibinfo
  {note} {pMID: 37358397}\BibitemShut {NoStop}%
\bibitem [{\citenamefont {Heinrich}\ \emph {et~al.}(2021)\citenamefont
  {Heinrich}, \citenamefont {Taucer}, \citenamefont {Kfir}, \citenamefont
  {Corkum}, \citenamefont {Staudte}, \citenamefont {Ropers},\ and\
  \citenamefont {Sivis}}]{chiral}%
  \BibitemOpen
  \bibfield  {author} {\bibinfo {author} {\bibfnamefont {T.}~\bibnamefont
  {Heinrich}}, \bibinfo {author} {\bibfnamefont {M.}~\bibnamefont {Taucer}},
  \bibinfo {author} {\bibfnamefont {O.}~\bibnamefont {Kfir}}, \bibinfo {author}
  {\bibfnamefont {P.~B.}\ \bibnamefont {Corkum}}, \bibinfo {author}
  {\bibfnamefont {A.}~\bibnamefont {Staudte}}, \bibinfo {author} {\bibfnamefont
  {C.}~\bibnamefont {Ropers}},\ and\ \bibinfo {author} {\bibfnamefont
  {M.}~\bibnamefont {Sivis}},\ }\bibfield  {title} {\bibinfo {title} {Chiral
  high-harmonic generation and spectroscopy on solid surfaces using
  polarization-tailored strong fields},\ }\href
  {https://doi.org/10.1038/s41467-021-23999-9} {\bibfield  {journal} {\bibinfo
  {journal} {Nat Commun}\ }\textbf {\bibinfo {volume} {12}},\ \bibinfo {pages}
  {3723} (\bibinfo {year} {2021})}\BibitemShut {NoStop}%
\bibitem [{\citenamefont {Imai}\ \emph {et~al.}(2020)\citenamefont {Imai},
  \citenamefont {Ono},\ and\ \citenamefont {Ishihara}}]{correlations}%
  \BibitemOpen
  \bibfield  {author} {\bibinfo {author} {\bibfnamefont {S.}~\bibnamefont
  {Imai}}, \bibinfo {author} {\bibfnamefont {A.}~\bibnamefont {Ono}},\ and\
  \bibinfo {author} {\bibfnamefont {S.}~\bibnamefont {Ishihara}},\ }\bibfield
  {title} {\bibinfo {title} {High harmonic generation in a correlated electron
  system},\ }\href {https://doi.org/10.1103/PhysRevLett.124.157404} {\bibfield
  {journal} {\bibinfo  {journal} {Phys. Rev. Lett.}\ }\textbf {\bibinfo
  {volume} {124}},\ \bibinfo {pages} {157404} (\bibinfo {year}
  {2020})}\BibitemShut {NoStop}%
\bibitem [{\citenamefont {Scrinzi}(2014)}]{Scrinzi2014}%
  \BibitemOpen
  \bibfield  {author} {\bibinfo {author} {\bibfnamefont {A.}~\bibnamefont
  {Scrinzi}},\ }\bibfield  {title} {\bibinfo {title} {Time-dependent
  {S}chr\"odinger equation},\ }in\ \href@noop {} {\emph {\bibinfo {booktitle}
  {Attosecond and {XUV} Physics: Ultrafast Dynamics and Spectroscopy}}},\
  \bibinfo {editor} {edited by\ \bibinfo {editor} {\bibfnamefont
  {T.}~\bibnamefont {Schultz}}\ and\ \bibinfo {editor} {\bibfnamefont
  {M.}~\bibnamefont {Vrakking}}}\ (\bibinfo  {publisher} {Wiley-{VCH}},\
  \bibinfo {address} {Weinheim},\ \bibinfo {year} {2014})\ pp.\ \bibinfo
  {pages} {257--292}\BibitemShut {NoStop}%
\bibitem [{\citenamefont {Lewenstein}\ \emph {et~al.}(1994)\citenamefont
  {Lewenstein}, \citenamefont {Balcou}, \citenamefont {Ivanov}, \citenamefont
  {L'Huillier},\ and\ \citenamefont {Corkum}}]{Lewenstein1994}%
  \BibitemOpen
  \bibfield  {author} {\bibinfo {author} {\bibfnamefont {M.}~\bibnamefont
  {Lewenstein}}, \bibinfo {author} {\bibfnamefont {P.}~\bibnamefont {Balcou}},
  \bibinfo {author} {\bibfnamefont {M.~Y.}\ \bibnamefont {Ivanov}}, \bibinfo
  {author} {\bibfnamefont {A.}~\bibnamefont {L'Huillier}},\ and\ \bibinfo
  {author} {\bibfnamefont {P.~B.}\ \bibnamefont {Corkum}},\ }\bibfield  {title}
  {\bibinfo {title} {Theory of high-harmonic generation by low-frequency laser
  fields},\ }\href {https://doi.org/10.1103/PhysRevA.49.2117} {\bibfield
  {journal} {\bibinfo  {journal} {Phys. Rev. A}\ }\textbf {\bibinfo {volume}
  {49}},\ \bibinfo {pages} {2117} (\bibinfo {year} {1994})}\BibitemShut
  {NoStop}%
\bibitem [{\citenamefont {Amini}\ \emph {et~al.}(2019)\citenamefont {Amini},
  \citenamefont {Biegert}, \citenamefont {Calegari}, \citenamefont
  {Chac{\'o}n}, \citenamefont {Ciappina}, \citenamefont {Dauphin},
  \citenamefont {Efimov}, \citenamefont {Figueira~de Morisson~Faria},
  \citenamefont {Giergiel}, \citenamefont {Gniewek}, \citenamefont {Landsman},
  \citenamefont {Lesiuk}, \citenamefont {Mandrysz}, \citenamefont {Maxwell},
  \citenamefont {Moszy{\'n}ski}, \citenamefont {Ortmann}, \citenamefont
  {P{\'e}rez-Hern{\'a}ndez}, \citenamefont {Pic{\'o}n}, \citenamefont
  {Pisanty}, \citenamefont {Prauzner-Bechcicki}, \citenamefont {Sacha},
  \citenamefont {Su{\'a}rez}, \citenamefont {Za{\"i}r}, \citenamefont
  {Zakrzewski},\ and\ \citenamefont {Lewenstein}}]{Amini2019}%
  \BibitemOpen
  \bibfield  {author} {\bibinfo {author} {\bibfnamefont {K.}~\bibnamefont
  {Amini}}, \bibinfo {author} {\bibfnamefont {J.}~\bibnamefont {Biegert}},
  \bibinfo {author} {\bibfnamefont {F.}~\bibnamefont {Calegari}}, \bibinfo
  {author} {\bibfnamefont {A.}~\bibnamefont {Chac{\'o}n}}, \bibinfo {author}
  {\bibfnamefont {M.~F.}\ \bibnamefont {Ciappina}}, \bibinfo {author}
  {\bibfnamefont {A.}~\bibnamefont {Dauphin}}, \bibinfo {author} {\bibfnamefont
  {D.~K.}\ \bibnamefont {Efimov}}, \bibinfo {author} {\bibfnamefont
  {C.}~\bibnamefont {Figueira~de Morisson~Faria}}, \bibinfo {author}
  {\bibfnamefont {K.}~\bibnamefont {Giergiel}}, \bibinfo {author}
  {\bibfnamefont {P.}~\bibnamefont {Gniewek}}, \bibinfo {author} {\bibfnamefont
  {A.~S.}\ \bibnamefont {Landsman}}, \bibinfo {author} {\bibfnamefont
  {M.}~\bibnamefont {Lesiuk}}, \bibinfo {author} {\bibfnamefont
  {M.}~\bibnamefont {Mandrysz}}, \bibinfo {author} {\bibfnamefont {A.~S.}\
  \bibnamefont {Maxwell}}, \bibinfo {author} {\bibfnamefont {R.}~\bibnamefont
  {Moszy{\'n}ski}}, \bibinfo {author} {\bibfnamefont {L.}~\bibnamefont
  {Ortmann}}, \bibinfo {author} {\bibfnamefont {J.~A.}\ \bibnamefont
  {P{\'e}rez-Hern{\'a}ndez}}, \bibinfo {author} {\bibfnamefont
  {A.}~\bibnamefont {Pic{\'o}n}}, \bibinfo {author} {\bibfnamefont
  {E.}~\bibnamefont {Pisanty}}, \bibinfo {author} {\bibfnamefont
  {J.}~\bibnamefont {Prauzner-Bechcicki}}, \bibinfo {author} {\bibfnamefont
  {K.}~\bibnamefont {Sacha}}, \bibinfo {author} {\bibfnamefont
  {N.}~\bibnamefont {Su{\'a}rez}}, \bibinfo {author} {\bibfnamefont
  {A.}~\bibnamefont {Za{\"i}r}}, \bibinfo {author} {\bibfnamefont
  {J.}~\bibnamefont {Zakrzewski}},\ and\ \bibinfo {author} {\bibfnamefont
  {M.}~\bibnamefont {Lewenstein}},\ }\bibfield  {title} {\bibinfo {title}
  {Symphony on strong field approximation},\ }\href
  {https://doi.org/10.1088/1361-6633/ab2bb1} {\bibfield  {journal} {\bibinfo
  {journal} {Rep. Progr. Phys.}\ }\textbf {\bibinfo {volume} {82}},\ \bibinfo
  {pages} {116001} (\bibinfo {year} {2019})}\BibitemShut {NoStop}%
\bibitem [{\citenamefont {Granados}\ and\ \citenamefont
  {Plaja}(2014)}]{Camilo}%
  \BibitemOpen
  \bibfield  {author} {\bibinfo {author} {\bibfnamefont {C.}~\bibnamefont
  {Granados}}\ and\ \bibinfo {author} {\bibfnamefont {L.}~\bibnamefont
  {Plaja}},\ }\bibfield  {title} {\bibinfo {title} {Space-time description of
  strong-field ionization and high-order-harmonic generation},\ }\href
  {https://doi.org/10.1103/PhysRevA.89.023428} {\bibfield  {journal} {\bibinfo
  {journal} {Phys. Rev. A}\ }\textbf {\bibinfo {volume} {89}},\ \bibinfo
  {pages} {023428} (\bibinfo {year} {2014})}\BibitemShut {NoStop}%
\bibitem [{\citenamefont {Sali{\`e}res}\ \emph {et~al.}(2001)\citenamefont
  {Sali{\`e}res}, \citenamefont {Carr{\'e}}, \citenamefont {Le~D{\'e}roff},
  \citenamefont {Grasbon}, \citenamefont {Paulus}, \citenamefont {Walther},
  \citenamefont {Kopold}, \citenamefont {Becker}, \citenamefont
  {Milo{\v{s}}evi{\'c}}, \citenamefont {Sanpera},\ and\ \citenamefont
  {Lewenstein}}]{Salieres2001}%
  \BibitemOpen
  \bibfield  {author} {\bibinfo {author} {\bibfnamefont {P.}~\bibnamefont
  {Sali{\`e}res}}, \bibinfo {author} {\bibfnamefont {B.}~\bibnamefont
  {Carr{\'e}}}, \bibinfo {author} {\bibfnamefont {L.}~\bibnamefont
  {Le~D{\'e}roff}}, \bibinfo {author} {\bibfnamefont {F.}~\bibnamefont
  {Grasbon}}, \bibinfo {author} {\bibfnamefont {G.}~\bibnamefont {Paulus}},
  \bibinfo {author} {\bibfnamefont {H.}~\bibnamefont {Walther}}, \bibinfo
  {author} {\bibfnamefont {R.}~\bibnamefont {Kopold}}, \bibinfo {author}
  {\bibfnamefont {W.}~\bibnamefont {Becker}}, \bibinfo {author} {\bibfnamefont
  {D.}~\bibnamefont {Milo{\v{s}}evi{\'c}}}, \bibinfo {author} {\bibfnamefont
  {A.}~\bibnamefont {Sanpera}},\ and\ \bibinfo {author} {\bibfnamefont
  {M.}~\bibnamefont {Lewenstein}},\ }\bibfield  {title} {\bibinfo {title}
  {Feynman's path-integral approach for intense-laser-atom interactions},\
  }\href {https://doi.org/10.1126/science.108836} {\bibfield  {journal}
  {\bibinfo  {journal} {Science}\ }\textbf {\bibinfo {volume} {292}},\ \bibinfo
  {pages} {902} (\bibinfo {year} {2001})}\BibitemShut {NoStop}%
\bibitem [{\citenamefont {Paulus}\ \emph {et~al.}(2000)\citenamefont {Paulus},
  \citenamefont {Grasbon}, \citenamefont {Dreischuh}, \citenamefont {Walther},
  \citenamefont {Kopold},\ and\ \citenamefont {Becker}}]{Paulus2000}%
  \BibitemOpen
  \bibfield  {author} {\bibinfo {author} {\bibfnamefont {G.~G.}\ \bibnamefont
  {Paulus}}, \bibinfo {author} {\bibfnamefont {F.}~\bibnamefont {Grasbon}},
  \bibinfo {author} {\bibfnamefont {A.}~\bibnamefont {Dreischuh}}, \bibinfo
  {author} {\bibfnamefont {H.}~\bibnamefont {Walther}}, \bibinfo {author}
  {\bibfnamefont {R.}~\bibnamefont {Kopold}},\ and\ \bibinfo {author}
  {\bibfnamefont {W.}~\bibnamefont {Becker}},\ }\bibfield  {title} {\bibinfo
  {title} {Above-threshold ionization by an elliptically polarized field:
  Interplay between electronic quantum trajectories},\ }\href
  {https://doi.org/10.1103/PhysRevLett.84.3791} {\bibfield  {journal} {\bibinfo
   {journal} {Phys. Rev. Lett.}\ }\textbf {\bibinfo {volume} {84}},\ \bibinfo
  {pages} {3791} (\bibinfo {year} {2000})}\BibitemShut {NoStop}%
\bibitem [{\citenamefont {Kopold}\ \emph {et~al.}(2002)\citenamefont {Kopold},
  \citenamefont {Becker},\ and\ \citenamefont
  {Milo{\v{s}}evi{\'c}}}]{Kopold2002}%
  \BibitemOpen
  \bibfield  {author} {\bibinfo {author} {\bibfnamefont {R.}~\bibnamefont
  {Kopold}}, \bibinfo {author} {\bibfnamefont {W.}~\bibnamefont {Becker}},\
  and\ \bibinfo {author} {\bibfnamefont {D.~B.}\ \bibnamefont
  {Milo{\v{s}}evi{\'c}}},\ }\bibfield  {title} {\bibinfo {title} {Quantum
  orbits: a space-time picture of intense-laser-induced processes in atoms},\
  }\href {https://doi.org/10.1080/09500340210140731} {\bibfield  {journal}
  {\bibinfo  {journal} {J. Mod. Opt.}\ }\textbf {\bibinfo {volume} {49}},\
  \bibinfo {pages} {1987} (\bibinfo {year} {2002})}\BibitemShut {NoStop}%
\bibitem [{\citenamefont {Ivanov}\ and\ \citenamefont
  {Smirnova}(2014)}]{Ivanov2014}%
  \BibitemOpen
  \bibfield  {author} {\bibinfo {author} {\bibfnamefont {M.}~\bibnamefont
  {Ivanov}}\ and\ \bibinfo {author} {\bibfnamefont {O.}~\bibnamefont
  {Smirnova}},\ }\bibfield  {title} {\bibinfo {title} {Multielectron high
  harmonic generation: simple man on a complex plane},\ }in\ \href@noop {}
  {\emph {\bibinfo {booktitle} {Attosecond and {XUV} Physics: Ultrafast
  Dynamics and Spectroscopy}}},\ \bibinfo {editor} {edited by\ \bibinfo
  {editor} {\bibfnamefont {T.}~\bibnamefont {Schultz}}\ and\ \bibinfo {editor}
  {\bibfnamefont {M.}~\bibnamefont {Vrakking}}}\ (\bibinfo  {publisher}
  {Wiley-{VCH}},\ \bibinfo {address} {Weinheim},\ \bibinfo {year} {2014})\ pp.\
  \bibinfo {pages} {201--256}\BibitemShut {NoStop}%
\bibitem [{\citenamefont {Nayak}\ \emph {et~al.}(2019)\citenamefont {Nayak},
  \citenamefont {Dumergue}, \citenamefont {K{\"u}hn}, \citenamefont {Mondal},
  \citenamefont {Csizmadia}, \citenamefont {Harshitha}, \citenamefont
  {F{\"u}le}, \citenamefont {Kahaly}, \citenamefont {Farkas}, \citenamefont
  {Major}, \citenamefont {Szaszk\'o-Bog\'ar}, \citenamefont {F\"oldi},
  \citenamefont {Majorosi}, \citenamefont {Tsatrafyllis}, \citenamefont
  {Skantzakis}, \citenamefont {Neori\v{c}i\'c}, \citenamefont {Shirozhan},
  \citenamefont {Vampa}, \citenamefont {Varjú}, \citenamefont {Tzallas},
  \citenamefont {Sansone}, \citenamefont {Charalambidis},\ and\ \citenamefont
  {Kahaly}}]{Nayak2019}%
  \BibitemOpen
  \bibfield  {author} {\bibinfo {author} {\bibfnamefont {A.}~\bibnamefont
  {Nayak}}, \bibinfo {author} {\bibfnamefont {M.}~\bibnamefont {Dumergue}},
  \bibinfo {author} {\bibfnamefont {S.}~\bibnamefont {K{\"u}hn}}, \bibinfo
  {author} {\bibfnamefont {S.}~\bibnamefont {Mondal}}, \bibinfo {author}
  {\bibfnamefont {T.}~\bibnamefont {Csizmadia}}, \bibinfo {author}
  {\bibfnamefont {N.}~\bibnamefont {Harshitha}}, \bibinfo {author}
  {\bibfnamefont {M.}~\bibnamefont {F{\"u}le}}, \bibinfo {author}
  {\bibfnamefont {M.~U.}\ \bibnamefont {Kahaly}}, \bibinfo {author}
  {\bibfnamefont {B.}~\bibnamefont {Farkas}}, \bibinfo {author} {\bibfnamefont
  {B.}~\bibnamefont {Major}}, \bibinfo {author} {\bibfnamefont
  {V.}~\bibnamefont {Szaszk\'o-Bog\'ar}}, \bibinfo {author} {\bibfnamefont
  {P.}~\bibnamefont {F\"oldi}}, \bibinfo {author} {\bibfnamefont
  {S.}~\bibnamefont {Majorosi}}, \bibinfo {author} {\bibfnamefont
  {N.}~\bibnamefont {Tsatrafyllis}}, \bibinfo {author} {\bibfnamefont
  {E.}~\bibnamefont {Skantzakis}}, \bibinfo {author} {\bibfnamefont
  {L.}~\bibnamefont {Neori\v{c}i\'c}}, \bibinfo {author} {\bibfnamefont
  {M.}~\bibnamefont {Shirozhan}}, \bibinfo {author} {\bibfnamefont
  {V.}~\bibnamefont {Vampa}}, \bibinfo {author} {\bibfnamefont
  {K.}~\bibnamefont {Varjú}}, \bibinfo {author} {\bibfnamefont
  {P.}~\bibnamefont {Tzallas}}, \bibinfo {author} {\bibfnamefont
  {G.}~\bibnamefont {Sansone}}, \bibinfo {author} {\bibfnamefont
  {D.}~\bibnamefont {Charalambidis}},\ and\ \bibinfo {author} {\bibfnamefont
  {S.}~\bibnamefont {Kahaly}},\ }\bibfield  {title} {\bibinfo {title} {Saddle
  point approaches in strong field physics and generation of attosecond
  pulses},\ }\href {https://doi.org/10.1016/j.physrep.2019.10.002} {\bibfield
  {journal} {\bibinfo  {journal} {Phys. Rep.}\ }\textbf {\bibinfo {volume}
  {833}},\ \bibinfo {pages} {1} (\bibinfo {year} {2019})}\BibitemShut {NoStop}%
\bibitem [{\citenamefont {Kondo}\ \emph {et~al.}(1996)\citenamefont {Kondo},
  \citenamefont {Kobayashi}, \citenamefont {Sagisaka}, \citenamefont
  {Nabekawa},\ and\ \citenamefont {Watanabe}}]{KONDO1996}%
  \BibitemOpen
  \bibfield  {author} {\bibinfo {author} {\bibfnamefont {K.}~\bibnamefont
  {Kondo}}, \bibinfo {author} {\bibfnamefont {Y.}~\bibnamefont {Kobayashi}},
  \bibinfo {author} {\bibfnamefont {A.}~\bibnamefont {Sagisaka}}, \bibinfo
  {author} {\bibfnamefont {Y.}~\bibnamefont {Nabekawa}},\ and\ \bibinfo
  {author} {\bibfnamefont {S.}~\bibnamefont {Watanabe}},\ }\bibfield  {title}
  {\bibinfo {title} {Tunneling ionization and harmonic generation in two-color
  fields},\ }\href {https://doi.org/10.1364/JOSAB.13.000424} {\bibfield
  {journal} {\bibinfo  {journal} {J. Opt. Soc. Am. B}\ }\textbf {\bibinfo
  {volume} {13}},\ \bibinfo {pages} {424} (\bibinfo {year} {1996})}\BibitemShut
  {NoStop}%
\bibitem [{\citenamefont {Chipperfield}\ \emph {et~al.}(2009)\citenamefont
  {Chipperfield}, \citenamefont {Robinson}, \citenamefont {Tisch},\ and\
  \citenamefont {Marangos}}]{Chipperfield2009}%
  \BibitemOpen
  \bibfield  {author} {\bibinfo {author} {\bibfnamefont {L.~E.}\ \bibnamefont
  {Chipperfield}}, \bibinfo {author} {\bibfnamefont {J.~S.}\ \bibnamefont
  {Robinson}}, \bibinfo {author} {\bibfnamefont {J.~W.~G.}\ \bibnamefont
  {Tisch}},\ and\ \bibinfo {author} {\bibfnamefont {J.~P.}\ \bibnamefont
  {Marangos}},\ }\bibfield  {title} {\bibinfo {title} {Ideal waveform to
  generate the maximum possible electron recollision energy for any given
  oscillation period},\ }\href {https://doi.org/10.1103/PhysRevLett.102.063003}
  {\bibfield  {journal} {\bibinfo  {journal} {Phys. Rev. Lett.}\ }\textbf
  {\bibinfo {volume} {102}},\ \bibinfo {pages} {063003} (\bibinfo {year}
  {2009})}\BibitemShut {NoStop}%
\bibitem [{\citenamefont {Kim}\ \emph {et~al.}(2005)\citenamefont {Kim},
  \citenamefont {Kim}, \citenamefont {Kim}, \citenamefont {Lee}, \citenamefont
  {Lee}, \citenamefont {Park}, \citenamefont {Cho},\ and\ \citenamefont
  {Nam}}]{KIM2005}%
  \BibitemOpen
  \bibfield  {author} {\bibinfo {author} {\bibfnamefont {I.~J.}\ \bibnamefont
  {Kim}}, \bibinfo {author} {\bibfnamefont {C.~M.}\ \bibnamefont {Kim}},
  \bibinfo {author} {\bibfnamefont {H.~T.}\ \bibnamefont {Kim}}, \bibinfo
  {author} {\bibfnamefont {G.~H.}\ \bibnamefont {Lee}}, \bibinfo {author}
  {\bibfnamefont {Y.~S.}\ \bibnamefont {Lee}}, \bibinfo {author} {\bibfnamefont
  {J.~Y.}\ \bibnamefont {Park}}, \bibinfo {author} {\bibfnamefont {D.~J.}\
  \bibnamefont {Cho}},\ and\ \bibinfo {author} {\bibfnamefont {C.~H.}\
  \bibnamefont {Nam}},\ }\bibfield  {title} {\bibinfo {title} {Highly efficient
  high-harmonic generation in an orthogonally polarized two-color laser
  field},\ }\href {https://doi.org/10.1103/PhysRevLett.94.243901} {\bibfield
  {journal} {\bibinfo  {journal} {Phys. Rev. Lett.}\ }\textbf {\bibinfo
  {volume} {94}},\ \bibinfo {pages} {243901} (\bibinfo {year}
  {2005})}\BibitemShut {NoStop}%
\bibitem [{\citenamefont {Mauritsson}\ \emph {et~al.}(2006)\citenamefont
  {Mauritsson}, \citenamefont {Johnsson}, \citenamefont {Gustafsson},
  \citenamefont {L'Huillier}, \citenamefont {Schafer},\ and\ \citenamefont
  {Gaarde}}]{MAURITSSON2006}%
  \BibitemOpen
  \bibfield  {author} {\bibinfo {author} {\bibfnamefont {J.}~\bibnamefont
  {Mauritsson}}, \bibinfo {author} {\bibfnamefont {P.}~\bibnamefont
  {Johnsson}}, \bibinfo {author} {\bibfnamefont {E.}~\bibnamefont
  {Gustafsson}}, \bibinfo {author} {\bibfnamefont {A.}~\bibnamefont
  {L'Huillier}}, \bibinfo {author} {\bibfnamefont {K.~J.}\ \bibnamefont
  {Schafer}},\ and\ \bibinfo {author} {\bibfnamefont {M.~B.}\ \bibnamefont
  {Gaarde}},\ }\bibfield  {title} {\bibinfo {title} {Attosecond pulse trains
  generated using two color laser fields},\ }\href
  {https://doi.org/10.1103/PhysRevLett.97.013001} {\bibfield  {journal}
  {\bibinfo  {journal} {Phys. Rev. Lett.}\ }\textbf {\bibinfo {volume} {97}},\
  \bibinfo {pages} {013001} (\bibinfo {year} {2006})}\BibitemShut {NoStop}%
\bibitem [{\citenamefont {Brizuela}\ \emph {et~al.}(2013)\citenamefont
  {Brizuela}, \citenamefont {Heyl}, \citenamefont {Rudawski}, \citenamefont
  {Kroon}, \citenamefont {Rading}, \citenamefont {Dahlstr\"om}, \citenamefont
  {Mauritsson}, \citenamefont {Johnsson}, \citenamefont {Arnold},\ and\
  \citenamefont {L'Huillier}}]{HUILLIER2013}%
  \BibitemOpen
  \bibfield  {author} {\bibinfo {author} {\bibfnamefont {F.}~\bibnamefont
  {Brizuela}}, \bibinfo {author} {\bibfnamefont {C.~M.}\ \bibnamefont {Heyl}},
  \bibinfo {author} {\bibfnamefont {P.}~\bibnamefont {Rudawski}}, \bibinfo
  {author} {\bibfnamefont {D.}~\bibnamefont {Kroon}}, \bibinfo {author}
  {\bibfnamefont {L.}~\bibnamefont {Rading}}, \bibinfo {author} {\bibfnamefont
  {J.~M.}\ \bibnamefont {Dahlstr\"om}}, \bibinfo {author} {\bibfnamefont
  {J.}~\bibnamefont {Mauritsson}}, \bibinfo {author} {\bibfnamefont
  {P.}~\bibnamefont {Johnsson}}, \bibinfo {author} {\bibfnamefont {C.~L.}\
  \bibnamefont {Arnold}},\ and\ \bibinfo {author} {\bibfnamefont
  {A.}~\bibnamefont {L'Huillier}},\ }\bibfield  {title} {\bibinfo {title}
  {Efficient high-order harmonic generation boosted by below-threshold
  harmonics},\ }\href@noop {} {\bibfield  {journal} {\bibinfo  {journal} {Sci.
  Rep.}\ }\textbf {\bibinfo {volume} {3}},\ \bibinfo {pages} {1410} (\bibinfo
  {year} {2013})}\BibitemShut {NoStop}%
\bibitem [{\citenamefont {Smirnova}\ \emph {et~al.}(2009)\citenamefont
  {Smirnova}, \citenamefont {Mairesse}, \citenamefont {Patchkovskii},
  \citenamefont {Dudovich}, \citenamefont {Villeneuve}, \citenamefont
  {Corkum},\ and\ \citenamefont {Ivanov}}]{IVANOV_2008}%
  \BibitemOpen
  \bibfield  {author} {\bibinfo {author} {\bibfnamefont {O.}~\bibnamefont
  {Smirnova}}, \bibinfo {author} {\bibfnamefont {Y.}~\bibnamefont {Mairesse}},
  \bibinfo {author} {\bibfnamefont {S.}~\bibnamefont {Patchkovskii}}, \bibinfo
  {author} {\bibfnamefont {N.}~\bibnamefont {Dudovich}}, \bibinfo {author}
  {\bibfnamefont {D.}~\bibnamefont {Villeneuve}}, \bibinfo {author}
  {\bibfnamefont {P.}~\bibnamefont {Corkum}},\ and\ \bibinfo {author}
  {\bibfnamefont {M.~Y.}\ \bibnamefont {Ivanov}},\ }\bibfield  {title}
  {\bibinfo {title} {High harmonic interferometry of multi-electron dynamics in
  molecules},\ }\href {https://doi.org/https://doi.org/10.1038/nature08253}
  {\bibfield  {journal} {\bibinfo  {journal} {Nature}\ }\textbf {\bibinfo
  {volume} {460}},\ \bibinfo {pages} {972} (\bibinfo {year}
  {2009})}\BibitemShut {NoStop}%
\bibitem [{\citenamefont {Vampa}\ \emph {et~al.}(2015)\citenamefont {Vampa},
  \citenamefont {Hammond}, \citenamefont {Thir\'e}, \citenamefont {Schmidt},
  \citenamefont {L\'egar\'e}, \citenamefont {McDonald}, \citenamefont {Brabec},
  \citenamefont {Klug},\ and\ \citenamefont {Corkum}}]{CORKUM_2015}%
  \BibitemOpen
  \bibfield  {author} {\bibinfo {author} {\bibfnamefont {G.}~\bibnamefont
  {Vampa}}, \bibinfo {author} {\bibfnamefont {T.~J.}\ \bibnamefont {Hammond}},
  \bibinfo {author} {\bibfnamefont {N.}~\bibnamefont {Thir\'e}}, \bibinfo
  {author} {\bibfnamefont {B.~E.}\ \bibnamefont {Schmidt}}, \bibinfo {author}
  {\bibfnamefont {F.}~\bibnamefont {L\'egar\'e}}, \bibinfo {author}
  {\bibfnamefont {C.~R.}\ \bibnamefont {McDonald}}, \bibinfo {author}
  {\bibfnamefont {T.}~\bibnamefont {Brabec}}, \bibinfo {author} {\bibfnamefont
  {D.~D.}\ \bibnamefont {Klug}},\ and\ \bibinfo {author} {\bibfnamefont
  {P.~B.}\ \bibnamefont {Corkum}},\ }\bibfield  {title} {\bibinfo {title}
  {All-optical reconstruction of crystal band structure},\ }\href
  {https://doi.org/10.1103/PhysRevLett.115.193603} {\bibfield  {journal}
  {\bibinfo  {journal} {Phys. Rev. Lett.}\ }\textbf {\bibinfo {volume} {115}},\
  \bibinfo {pages} {193603} (\bibinfo {year} {2015})}\BibitemShut {NoStop}%
\bibitem [{\citenamefont {Uzan-Narovlansky}\ \emph {et~al.}(2022)\citenamefont
  {Uzan-Narovlansky}, \citenamefont {Jim\'enez-Gasl\'an}, \citenamefont
  {Orenstein}, \citenamefont {Silva}, \citenamefont {Arusi-Parpar},
  \citenamefont {Shames}, \citenamefont {Bruner}, \citenamefont {Yan},
  \citenamefont {Smirnova}, \citenamefont {Ivanov},\ and\ \citenamefont
  {Dudovich}}]{DUDOVICH_2022}%
  \BibitemOpen
  \bibfield  {author} {\bibinfo {author} {\bibfnamefont {A.~J.}\ \bibnamefont
  {Uzan-Narovlansky}}, \bibinfo {author} {\bibfnamefont {A.}~\bibnamefont
  {Jim\'enez-Gasl\'an}}, \bibinfo {author} {\bibfnamefont {G.}~\bibnamefont
  {Orenstein}}, \bibinfo {author} {\bibfnamefont {R.~E.~F.}\ \bibnamefont
  {Silva}}, \bibinfo {author} {\bibfnamefont {T.}~\bibnamefont {Arusi-Parpar}},
  \bibinfo {author} {\bibfnamefont {S.}~\bibnamefont {Shames}}, \bibinfo
  {author} {\bibfnamefont {B.~D.}\ \bibnamefont {Bruner}}, \bibinfo {author}
  {\bibfnamefont {B.}~\bibnamefont {Yan}}, \bibinfo {author} {\bibfnamefont
  {O.}~\bibnamefont {Smirnova}}, \bibinfo {author} {\bibfnamefont
  {M.}~\bibnamefont {Ivanov}},\ and\ \bibinfo {author} {\bibfnamefont
  {N.}~\bibnamefont {Dudovich}},\ }\bibfield  {title} {\bibinfo {title}
  {Observation of light-driven band structure via multiband high-harmonic
  spectroscopy},\ }\href
  {https://doi.org/https://doi.org/10.1038/s41566-022-01010-1} {\bibfield
  {journal} {\bibinfo  {journal} {Nat. Photon.}\ }\textbf {\bibinfo {volume}
  {16}},\ \bibinfo {pages} {428} (\bibinfo {year} {2022})}\BibitemShut
  {NoStop}%
\bibitem [{\citenamefont {Tekavec}\ \emph {et~al.}(2007)\citenamefont
  {Tekavec}, \citenamefont {Lott},\ and\ \citenamefont {Marcus}}]{MARCUS2007}%
  \BibitemOpen
  \bibfield  {author} {\bibinfo {author} {\bibfnamefont {P.~F.}\ \bibnamefont
  {Tekavec}}, \bibinfo {author} {\bibfnamefont {G.~A.}\ \bibnamefont {Lott}},\
  and\ \bibinfo {author} {\bibfnamefont {A.~H.}\ \bibnamefont {Marcus}},\
  }\bibfield  {title} {\bibinfo {title} {Fluorescence-detected two-dimensional
  electronic coherence spectroscopy by acousto-optic phase modulation},\
  }\href@noop {} {\bibfield  {journal} {\bibinfo  {journal} {J. Chem. Phys.}\
  }\textbf {\bibinfo {volume} {127}},\ \bibinfo {pages} {214307} (\bibinfo
  {year} {2007})}\BibitemShut {NoStop}%
\bibitem [{\citenamefont {Karki}\ \emph {et~al.}(2014)\citenamefont {Karki},
  \citenamefont {Widom}, \citenamefont {Seibt}, \citenamefont {Moody},
  \citenamefont {Lonergan}, \citenamefont {onu Pullerits},\ and\ \citenamefont
  {Marcus}}]{KARKI2014C}%
  \BibitemOpen
  \bibfield  {author} {\bibinfo {author} {\bibfnamefont {K.~J.}\ \bibnamefont
  {Karki}}, \bibinfo {author} {\bibfnamefont {J.~R.}\ \bibnamefont {Widom}},
  \bibinfo {author} {\bibfnamefont {J.}~\bibnamefont {Seibt}}, \bibinfo
  {author} {\bibfnamefont {I.}~\bibnamefont {Moody}}, \bibinfo {author}
  {\bibfnamefont {M.~C.}\ \bibnamefont {Lonergan}}, \bibinfo {author}
  {\bibfnamefont {T.}~\bibnamefont {onu Pullerits}},\ and\ \bibinfo {author}
  {\bibfnamefont {A.~H.}\ \bibnamefont {Marcus}},\ }\bibfield  {title}
  {\bibinfo {title} {Coherent two-dimensional photocurrent spectroscopy in a
  pbs quantum dot photocell},\ }\href@noop {} {\bibfield  {journal} {\bibinfo
  {journal} {Nat. Commun.}\ }\textbf {\bibinfo {volume} {5}} (\bibinfo {year}
  {2014})}\BibitemShut {NoStop}%
\bibitem [{\citenamefont {Wagner}\ \emph {et~al.}(2005)\citenamefont {Wagner},
  \citenamefont {Li}, \citenamefont {Semmlow},\ and\ \citenamefont
  {Warren}}]{WARREN_2005}%
  \BibitemOpen
  \bibfield  {author} {\bibinfo {author} {\bibfnamefont {W.}~\bibnamefont
  {Wagner}}, \bibinfo {author} {\bibfnamefont {C.}~\bibnamefont {Li}}, \bibinfo
  {author} {\bibfnamefont {J.}~\bibnamefont {Semmlow}},\ and\ \bibinfo {author}
  {\bibfnamefont {W.~S.}\ \bibnamefont {Warren}},\ }\bibfield  {title}
  {\bibinfo {title} {Rapid phase-cycled two-dimensional optical spectroscopy in
  fluorescence and transmission mode},\ }\href@noop {} {\bibfield  {journal}
  {\bibinfo  {journal} {Opt. Express}\ }\textbf {\bibinfo {volume} {13}},\
  \bibinfo {pages} {3697} (\bibinfo {year} {2005})}\BibitemShut {NoStop}%
\bibitem [{\citenamefont {Karki}\ \emph {et~al.}(2019)\citenamefont {Karki},
  \citenamefont {Chen}, \citenamefont {Sakurai}, \citenamefont {Shi},
  \citenamefont {Gardiner}, \citenamefont {Kühn}, \citenamefont {Cogdell},\
  and\ \citenamefont {Pullerits}}]{KARKI_2019C}%
  \BibitemOpen
  \bibfield  {author} {\bibinfo {author} {\bibfnamefont {K.~J.}\ \bibnamefont
  {Karki}}, \bibinfo {author} {\bibfnamefont {J.}~\bibnamefont {Chen}},
  \bibinfo {author} {\bibfnamefont {A.}~\bibnamefont {Sakurai}}, \bibinfo
  {author} {\bibfnamefont {Q.}~\bibnamefont {Shi}}, \bibinfo {author}
  {\bibfnamefont {A.~T.}\ \bibnamefont {Gardiner}}, \bibinfo {author}
  {\bibfnamefont {O.}~\bibnamefont {Kühn}}, \bibinfo {author} {\bibfnamefont
  {R.~J.}\ \bibnamefont {Cogdell}},\ and\ \bibinfo {author} {\bibfnamefont
  {T.}~\bibnamefont {Pullerits}},\ }\bibfield  {title} {\bibinfo {title}
  {Before förster. initial excitation in photosynthetic light harvesting},\
  }\href {https://doi.org/10.1039/C9SC01888C} {\bibfield  {journal} {\bibinfo
  {journal} {Chem. Sci.}\ }\textbf {\bibinfo {volume} {10}},\ \bibinfo {pages}
  {7923} (\bibinfo {year} {2019})}\BibitemShut {NoStop}%
\bibitem [{\citenamefont {Damtie}\ \emph {et~al.}(2017)\citenamefont {Damtie},
  \citenamefont {Wacker}, \citenamefont {Pullerits},\ and\ \citenamefont
  {Karki}}]{KARKI_2017C}%
  \BibitemOpen
  \bibfield  {author} {\bibinfo {author} {\bibfnamefont {F.~A.}\ \bibnamefont
  {Damtie}}, \bibinfo {author} {\bibfnamefont {A.}~\bibnamefont {Wacker}},
  \bibinfo {author} {\bibfnamefont {T.}~\bibnamefont {Pullerits}},\ and\
  \bibinfo {author} {\bibfnamefont {K.~J.}\ \bibnamefont {Karki}},\ }\bibfield
  {title} {\bibinfo {title} {Two-dimensional action spectroscopy of excitonic
  systems: Explicit simulation using a phase-modulation technique},\
  }\href@noop {} {\bibfield  {journal} {\bibinfo  {journal} {Phys. Rev. A}\
  }\textbf {\bibinfo {volume} {96}},\ \bibinfo {pages} {053830} (\bibinfo
  {year} {2017})}\BibitemShut {NoStop}%
\bibitem [{\citenamefont {Kalaee}\ \emph {et~al.}(2019)\citenamefont {Kalaee},
  \citenamefont {Damtie},\ and\ \citenamefont {Karki}}]{KARKI_2019B}%
  \BibitemOpen
  \bibfield  {author} {\bibinfo {author} {\bibfnamefont {A.~A.~S.}\
  \bibnamefont {Kalaee}}, \bibinfo {author} {\bibfnamefont {F.}~\bibnamefont
  {Damtie}},\ and\ \bibinfo {author} {\bibfnamefont {K.~J.}\ \bibnamefont
  {Karki}},\ }\bibfield  {title} {\bibinfo {title} {Differentiation of true
  nonlinear and incoherent mixing of linear signals in action-detected 2d
  spectroscopy},\ }\href {https://doi.org/10.1021/acs.jpca.9b01129} {\bibfield
  {journal} {\bibinfo  {journal} {J. Phys. Chem. A}\ }\textbf {\bibinfo
  {volume} {123}},\ \bibinfo {pages} {4119} (\bibinfo {year}
  {2019})}\BibitemShut {NoStop}%
\bibitem [{\citenamefont {Anda}\ and\ \citenamefont {Cole}(2021)}]{ANDA_2021}%
  \BibitemOpen
  \bibfield  {author} {\bibinfo {author} {\bibfnamefont {A.}~\bibnamefont
  {Anda}}\ and\ \bibinfo {author} {\bibfnamefont {J.~H.}\ \bibnamefont
  {Cole}},\ }\bibfield  {title} {\bibinfo {title} {{Two-dimensional
  spectroscopy beyond the perturbative limit: The influence of finite pulses
  and detection modes}},\ }\href {https://doi.org/10.1063/5.0038550} {\bibfield
   {journal} {\bibinfo  {journal} {J. Chem. Phys.}\ }\textbf {\bibinfo {volume}
  {154}},\ \bibinfo {pages} {114113} (\bibinfo {year} {2021})}\BibitemShut
  {NoStop}%
\bibitem [{\citenamefont {Bruschi}\ \emph {et~al.}(2022)\citenamefont
  {Bruschi}, \citenamefont {Gallina},\ and\ \citenamefont
  {Fresch}}]{BRUSCHI_2022}%
  \BibitemOpen
  \bibfield  {author} {\bibinfo {author} {\bibfnamefont {M.}~\bibnamefont
  {Bruschi}}, \bibinfo {author} {\bibfnamefont {F.}~\bibnamefont {Gallina}},\
  and\ \bibinfo {author} {\bibfnamefont {B.}~\bibnamefont {Fresch}},\
  }\bibfield  {title} {\bibinfo {title} {Simulating action-2d electronic
  spectroscopy of quantum dots: insights on the exciton and biexciton interplay
  from detection-mode and time-gating},\ }\href
  {https://doi.org/10.1039/D2CP04270C} {\bibfield  {journal} {\bibinfo
  {journal} {Phys. Chem. Chem. Phys.}\ }\textbf {\bibinfo {volume} {24}},\
  \bibinfo {pages} {27645} (\bibinfo {year} {2022})}\BibitemShut {NoStop}%
\bibitem [{\citenamefont {Ciappina}\ \emph {et~al.}(2017)\citenamefont
  {Ciappina}, \citenamefont {Pérez-Hernández}, \citenamefont {Landsman},
  \citenamefont {Okell}, \citenamefont {Zherebtsov}, \citenamefont {Förg},
  \citenamefont {Schötz}, \citenamefont {Seiffert}, \citenamefont {Fennel},
  \citenamefont {Shaaran}, \citenamefont {Zimmermann}, \citenamefont {Chacón},
  \citenamefont {Guichard}, \citenamefont {Zaïr}, \citenamefont {Tisch},
  \citenamefont {Marangos}, \citenamefont {Witting}, \citenamefont {Braun},
  \citenamefont {Maier}, \citenamefont {Roso}, \citenamefont {Krüger},
  \citenamefont {Hommelhoff}, \citenamefont {Kling}, \citenamefont {Krausz},\
  and\ \citenamefont {Lewenstein}}]{Ciappina2017}%
  \BibitemOpen
  \bibfield  {author} {\bibinfo {author} {\bibfnamefont {M.~F.}\ \bibnamefont
  {Ciappina}}, \bibinfo {author} {\bibfnamefont {J.~A.}\ \bibnamefont
  {Pérez-Hernández}}, \bibinfo {author} {\bibfnamefont {A.~S.}\ \bibnamefont
  {Landsman}}, \bibinfo {author} {\bibfnamefont {W.~A.}\ \bibnamefont {Okell}},
  \bibinfo {author} {\bibfnamefont {S.}~\bibnamefont {Zherebtsov}}, \bibinfo
  {author} {\bibfnamefont {B.}~\bibnamefont {Förg}}, \bibinfo {author}
  {\bibfnamefont {J.}~\bibnamefont {Schötz}}, \bibinfo {author} {\bibfnamefont
  {L.}~\bibnamefont {Seiffert}}, \bibinfo {author} {\bibfnamefont
  {T.}~\bibnamefont {Fennel}}, \bibinfo {author} {\bibfnamefont
  {T.}~\bibnamefont {Shaaran}}, \bibinfo {author} {\bibfnamefont
  {T.}~\bibnamefont {Zimmermann}}, \bibinfo {author} {\bibfnamefont
  {A.}~\bibnamefont {Chacón}}, \bibinfo {author} {\bibfnamefont
  {R.}~\bibnamefont {Guichard}}, \bibinfo {author} {\bibfnamefont
  {A.}~\bibnamefont {Zaïr}}, \bibinfo {author} {\bibfnamefont {J.~W.~G.}\
  \bibnamefont {Tisch}}, \bibinfo {author} {\bibfnamefont {J.~P.}\ \bibnamefont
  {Marangos}}, \bibinfo {author} {\bibfnamefont {T.}~\bibnamefont {Witting}},
  \bibinfo {author} {\bibfnamefont {A.}~\bibnamefont {Braun}}, \bibinfo
  {author} {\bibfnamefont {S.~A.}\ \bibnamefont {Maier}}, \bibinfo {author}
  {\bibfnamefont {L.}~\bibnamefont {Roso}}, \bibinfo {author} {\bibfnamefont
  {M.}~\bibnamefont {Krüger}}, \bibinfo {author} {\bibfnamefont
  {P.}~\bibnamefont {Hommelhoff}}, \bibinfo {author} {\bibfnamefont {M.~F.}\
  \bibnamefont {Kling}}, \bibinfo {author} {\bibfnamefont {F.}~\bibnamefont
  {Krausz}},\ and\ \bibinfo {author} {\bibfnamefont {M.}~\bibnamefont
  {Lewenstein}},\ }\bibfield  {title} {\bibinfo {title} {Attosecond physics at
  the nanoscale},\ }\href {https://doi.org/10.1088/1361-6633/aa574e} {\bibfield
   {journal} {\bibinfo  {journal} {Rep. Progr. Phys.}\ }\textbf {\bibinfo
  {volume} {80}},\ \bibinfo {pages} {054401} (\bibinfo {year}
  {2017})}\BibitemShut {NoStop}%
\bibitem [{\citenamefont {Picca}\ \emph {et~al.}(2016)\citenamefont {Picca},
  \citenamefont {Gramajo},\ and\ \citenamefont {Garibotti}}]{Renata2016}%
  \BibitemOpen
  \bibfield  {author} {\bibinfo {author} {\bibfnamefont {R.~D.}\ \bibnamefont
  {Picca}}, \bibinfo {author} {\bibfnamefont {A.~A.}\ \bibnamefont {Gramajo}},\
  and\ \bibinfo {author} {\bibfnamefont {C.~R.}\ \bibnamefont {Garibotti}},\
  }\bibfield  {title} {\bibinfo {title} {Nonconstant ponderomotive energy in
  above-threshold ionization by intense short pulses},\ }\href
  {https://doi.org/10.1103/PhysRevA.93.023416} {\bibfield  {journal} {\bibinfo
  {journal} {Phys. Rev. A.}\ }\textbf {\bibinfo {volume} {93}},\ \bibinfo
  {pages} {023416} (\bibinfo {year} {2016})}\BibitemShut {NoStop}%
\bibitem [{\citenamefont {Milo{\v{s}}evi{\'c}}\ \emph
  {et~al.}(2016)\citenamefont {Milo{\v{s}}evi{\'c}}, \citenamefont {Paulus},
  \citenamefont {Bauer},\ and\ \citenamefont {Becker}}]{MilosevicReview}%
  \BibitemOpen
  \bibfield  {author} {\bibinfo {author} {\bibfnamefont {D.~B.}\ \bibnamefont
  {Milo{\v{s}}evi{\'c}}}, \bibinfo {author} {\bibfnamefont {G.}~\bibnamefont
  {Paulus}}, \bibinfo {author} {\bibfnamefont {D.}~\bibnamefont {Bauer}},\ and\
  \bibinfo {author} {\bibfnamefont {W.}~\bibnamefont {Becker}},\ }\bibfield
  {title} {\bibinfo {title} {Above-threshold ionization by few-cycle pulses},\
  }\href@noop {} {\bibfield  {journal} {\bibinfo  {journal} {J. Phys. B:\ At.
  Mol. Opt. Phys.}\ }\textbf {\bibinfo {volume} {39}},\ \bibinfo {pages} {R203}
  (\bibinfo {year} {2016})}\BibitemShut {NoStop}%
\end{thebibliography}%


\end{document}